%% file: ads2_v3.tex
\documentclass[11pt]{article}
\pdfoutput=1
\usepackage{latexsym}
\usepackage{chapterbib}
\usepackage{graphics,graphicx}
\usepackage{amssymb,amsmath,amsfonts,amstext}
\usepackage{hyperref}
\usepackage[bbgreekl]{mathbbol}
\usepackage{color,xcolor}
\usepackage{hyperref}
\usepackage{geometry}  
\usepackage{braket}
\usepackage{cancel}
\usepackage[utf8]{inputenc}

\include{macros}

\begin{document}

\begin{titlepage}

\pagestyle{empty}

\begin{flushright}
 {\small SISSA 44/2016/FISI\\ UPR-1281-T}
\end{flushright}
\vskip1.5in

\begin{center}
\textbf{\Large AdS$_2$ Holographic Dictionary}
\end{center}
\vskip0.2in

\begin{center}
{\large Mirjam Cveti\v c$^{a,b,}$\footnote{\href{mailto: cvetic@physics.upenn.edu}{\tt cvetic@physics.upenn.edu}}\thinspace ,   
Ioannis Papadimitriou$^{c,}$\footnote{\href{mailto: ipapadim@sissa.it}{\tt ioannis.papadimitriou@sissa.it}}}
\end{center}
\vskip0.2in

\begin{center}
{\small {$^{a}$}Department of Physics and Astronomy, University of Pennsylvania,\\ Philadelphia, PA 19104-6396, USA}\\ \vskip0.1in
{\small {$^{b}$}Center for Applied Mathematics and Theoretical Physics,\\
University of Maribor, SI2000 Maribor, Slovenia}\\ \vskip0.1in
{\small {$^{c}$}SISSA and INFN - Sezione di Trieste, Via Bonomea 265,\\
	34136 Trieste, Italy}
\end{center}
\vskip0.2in

\begin{abstract}

We construct the holographic dictionary for both running and constant dilaton solutions of the two dimensional Einstein-Maxwell-Dilaton theory that is obtained by a circle reduction from Einstein-Hilbert gravity with negative cosmological constant in three dimensions. This specific model ensures that the dual theory has a well defined ultraviolet completion in terms of a two dimensional conformal field theory, but our results apply qualitatively to a wider class of two dimensional dilaton gravity theories. For each type of solutions we perform holographic renormalization, compute the exact renormalized one-point functions in the presence of arbitrary sources, and derive the asymptotic symmetries and the corresponding conserved charges. In both cases we find that the scalar operator dual to the dilaton plays a crucial role in the description of the dynamics. Its source gives rise to a matter conformal anomaly for the running dilaton solutions, while its expectation value is the only non trivial observable for constant dilaton solutions. The role of this operator has been largely overlooked in the literature. We further show that the only non trivial conserved charges for running dilaton solutions are the mass and the electric charge, while for constant dilaton solutions only the electric charge is non zero. However, by uplifting the solutions to three dimensions we show that constant dilaton solutions can support non trivial extended symmetry algebras, including the one found by Comp\`ere, Song and Strominger \cite{Compere:2013bya}, in agreement with the results of Castro and Song \cite{Castro:2014ima}. Finally, we demonstrate that any solution of this specific dilaton gravity model can be uplifted to a family of asymptotically AdS$_2\times S^2$ or conformally AdS$_2\times S^2$ solutions of the STU model in four dimensions, including non extremal black holes. The four dimensional solutions obtained by uplifting the running dilaton solutions coincide with the so called `subtracted geometries', while those obtained from the uplift of the constant dilaton ones are new.

\end{abstract}

\end{titlepage}

\tableofcontents
\addtocontents{toc}{\protect\setcounter{tocdepth}{3}}

\renewcommand{\theequation}{\arabic{section}.\arabic{equation}}
\setcounter{page}{1} \pagestyle{plain}

\section{Introduction and summary of results}
\label{intro}
\setcounter{equation}{0}

Despite the plethora of gravity and string theory backgrounds that contain an AdS$_2$ region, arising for example in the near horizon limit of near extremal black holes \cite{Strominger:1998yg} or at the infrared of holographic renormalization group (RG) flows with finite charge density \cite{Chamblin:1999tk}, AdS$_2$ holography remains less understood than its higher dimensional cousins. Paradoxically, one of the main reasons is that it is apparently trivial: pure AdS$_2$ gravity does not allow finite energy excitations \cite{Maldacena:1998uz}. 

Nevertheless, AdS$_2$ holography has been studied extensively \cite{Strominger:1998yg,Cadoni:1999ja,Spradlin:1999bn,NavarroSalas:1999up,Cadoni:2000ah,Caldarelli:2000xk,Cadoni:2000gm,Hartman:2008dq,Alishahiha:2008tv,Castro:2008ms,Alishahiha:2008rt,Grumiller:2013swa,Almheiri:2014cka,Grumiller:2015vaa,Jensen:2016pah,Maldacena:2016upp,Engelsoy:2016xyb} and has been used to count the microstates of extremal black holes \cite{Sen:2008yk,Gupta:2008ki,Sen:2008vm}. Given the lack, until recently, of a good candidate for the holographic dual, the focus has been on attempts to describe the effects of the strong gravitational backreaction on AdS$_2$ by finite energy excitations. As elucidated recently by Almheiri and Polchinski \cite{Almheiri:2014cka} and further elaborated on in \cite{Maldacena:2016upp,Engelsoy:2016xyb}, to leading order the effect of the gravitational backreaction can be described by a rather universal AdS$_2$ dilaton gravity model. In \cite{Sachdev:2010um,Maldacena:2016hyu,Jensen:2016pah} it was argued that such a dilaton gravity model provides a holographic description of the infrared limit of the Sachdev-Ye-Kitaev model \cite{Sachdev1993,Kitaev}, a quantum mechanical system of Majorana fermions with random long range interactions. Moreover, AdS$_2$ dilaton gravity coupled to a gauge field can also provide a holographic description of the Kondo effect \cite{Erdmenger:2013dpa}.

In this paper we revisit the holographic dictionary and the asymptotic symmetries of  the specific 2D Einstein-Maxwell-Dilaton (EMD) model 
\bal\label{action}
S\sbtx{2D}= & \frac{1}{2\k_2^2}\(\int_\cm \text{d}^2\mathbf{x}\sqrt{-g}\;e^{-\j}\Big(R[g]+\frac{2}{L^2}-\frac14e^{-2\j}F_{ab}F^{ab}\Big)+\int_{\pa\cm} \text{d}t\sqrt{-\g}\;e^{-\j}2K\),
\eal
with the aim to clarify certain aspects of AdS$_2$ holography that have been to some extent elusive. The main motivation of our choice of model is that it can be obtained by a circle reduction from Einstein-Hilbert gravity with negative cosmological constant in three dimensions \cite{Strominger:1998yg}, which ensures that the dual theory admits a well defined ultraviolet (UV) completion in terms of a 2D CFT. Moreover, the model \eqref{action} arises as the very near horizon effective theory of nearly extremal black holes in five dimensions \cite{Strominger:1998yg}. Other studies of this specific model in the context of AdS$_2$ holography include \cite{Castro:2014ima,Almheiri:2016fws}. Similar models with a different coupling to the Maxwell field have been studied in \cite{Hartman:2008dq,Castro:2008ms,Grumiller:2015vaa,Jensen:2016pah}. Despite the fact that such models do not uplift to Einstein-Hilbert gravity in three dimensions, they share some qualitative properties with the model \eqref{action}, e.g., they generically admit two distinct classes of solutions, one with running dilaton, and one with constant dilaton. Clearly, the constant dilaton solutions coincide, but there are significant differences in the running dilaton solutions.   
Setting the Maxwell field in \eqref{action} consistently to zero results in the Jackiw–Teitelboim model \cite{Jackiw:1982hg,Teitelboim:1983fg}, which has been discussed recently in \cite{Almheiri:2014cka,Maldacena:2016upp,Engelsoy:2016xyb}. 

A second motivation for the EMD theory \eqref{action} is that it provides a holographic description of the so called `subtracted geometries' \cite{Cvetic:2011hp,Cvetic:2011dn,Cvetic:2012tr,Virmani:2012kw,Baggio:2012db,Cvetic:2013vqi,Cvetic:2014sxa}. These are asymptotically conformally AdS$_2\times S^2$ or AdS$_2\times S^3$ black holes that can be obtained through a `subtraction' procedure \cite{Cvetic:2011hp,Cvetic:2011dn} from generic multi-charge non extremal asymptotically flat black holes in four  \cite{Cvetic:1995kv,Cvetic:1996kv,Chong:2004na,Chow:2013tia,Chow:2014cca} and five \cite{Cvetic:1996xz} dimensions. More systematically, they can be obtained as a scaling limit \cite{Cvetic:2012tr}, or via Harrison transformations \cite{Virmani:2012kw,Cvetic:2013vqi}, but also through a {\em decoupling limit} where certain integration constants in the harmonic functions that describe the asymptotically flat {\em non extremal} black holes are set to zero \cite{Baggio:2012db}. The classical entropy of the subtracted geometries is the same as that of the original asymptotically flat black hole, but quantum corrections are different \cite{Cvetic:2014tka,Cvetic:2014eka,Cvetic:2015cca}.

\begin{figure}
	\scalebox{0.80}{\hskip-1.3in\includegraphics[height=3.8in, width=\textwidth, trim={0 0 0 1.4in},clip]{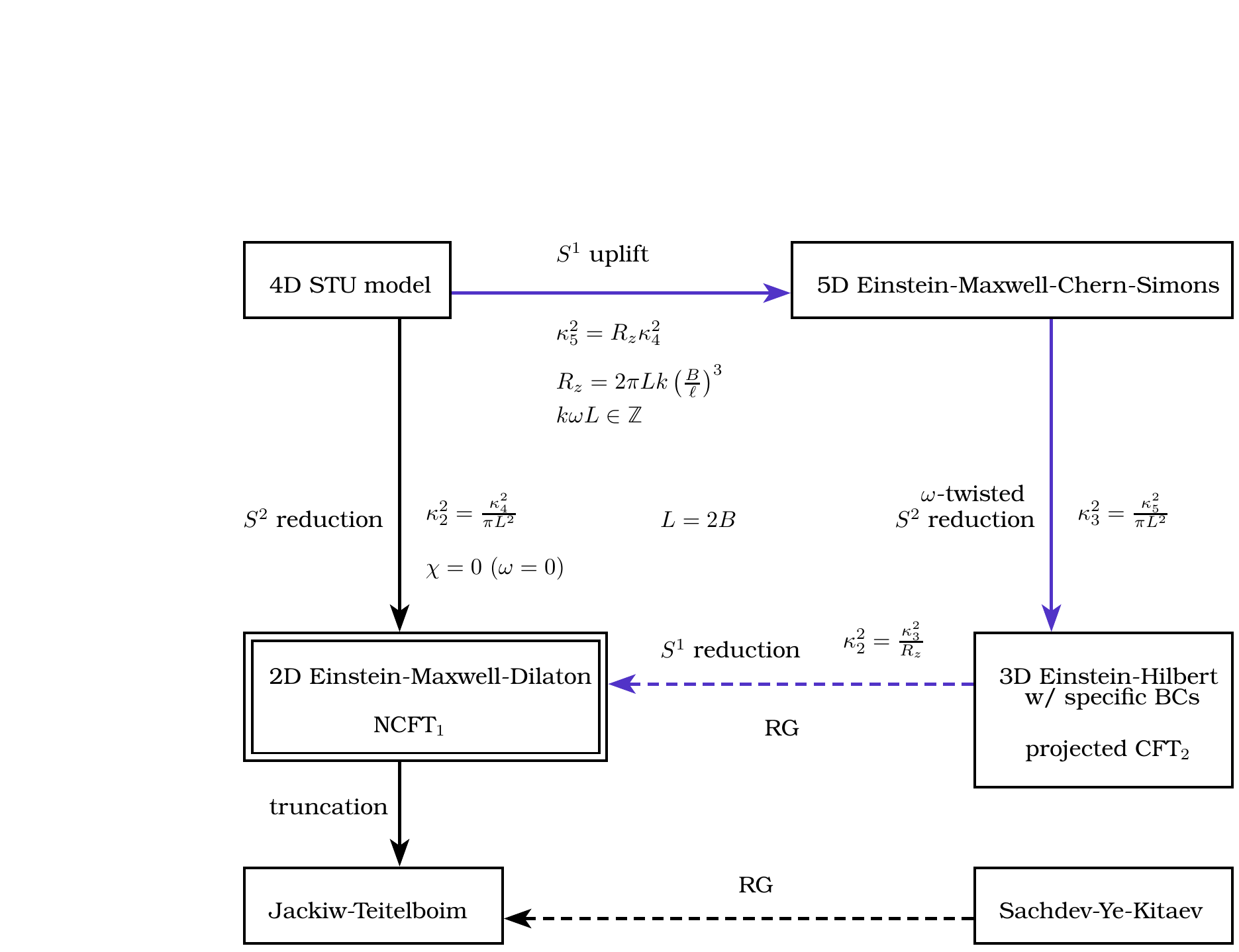}}
	\centering
	\caption{This diagram shows how the EMD model \eqref{action} arises as the low energy effective theory of non extremal asymptotically conformally AdS$_2\times S^2$ subtracted geometries in four dimensions, in the parameterization introduced in \cite{An:2016fzu}. For non rotating black holes the two routes to \eqref{action}, a direct $S^2$ reduction (black arrow) and the more general procedure through the uplift to five dimensions (blue arrows), coincide. However, only the latter is available for rotating black holes. The relevant Kaluza-Klein Ans\"atze are given in \eqref{KK-relations} and \eqref{KK-Ansatz}. A similar diagram applies for subtracted geometries in five dimensions, replacing $S^2$ with $S^3$.}
	\label{fig1}
\end{figure}

A holographic description of the asymptotically conformally AdS$_2\times S^2$ or AdS$_2\times S^3$ subtracted geometries requires a Kaluza-Klein  reduction on the compact manifold, in direct analogy with coincident D$p$ branes which are asymptotically conformal to AdS$_{p+2}\times S^{8-p}$ \cite{Itzhaki:1998dd,Kanitscheider:2008kd}. The 2D theory that is obtained by keeping only the massless modes coincides with \eqref{action}.
The relation with 3D gravity arises due to the fact that there is a linear dilaton that blows up in the ultraviolet, which forces one to consider instead the uplift of these black holes to five or six dimensions, where they become respectively
asymptotically AdS$_3\times S^2$ or AdS$_3\times S^3$ solutions \cite{Cvetic:2011hp,Cvetic:2011dn}. Kaluza-Klein reducing the uplifted solutions on their respective compact manifolds results in the uplift of the EMD theory \eqref{action} to Einstein-Hilbert gravity in three dimensions. The relation of the four dimensional subtracted geometries and their five dimensional uplift to Einstein Hilbert gravity in three dimensions and the AdS$_2$ theory \eqref{action} is depicted schematically in Fig. \ref{fig1}. The STU model in four dimensions and its subtracted geometries in the parameterization introduced in \cite{An:2016fzu} are summarized in appendix \ref{appendix}, together with the relevant Kaluza-Klein Ans\"atze.

\paragraph{Summary of results} Besides the general solutions for the 2D model \eqref{action} and their uplift to four dimensions, our main results relate to the holographic interpretation of these solutions, as well as the asymptotic symmetry algebras they support. Our findings partially agree with earlier studies, but there are also some significant differences, which we explain in detail.  

We find that the action \eqref{action} admits two distinct types of solutions, those with running dilaton and those with constant dilaton. Both are asymptotically AdS$_2$, but the AdS radius of the constant dilaton solutions is half that of the running dilaton ones. Moreover, constant dilaton solutions only exist in the presence of a non zero electric charge, while the role of the gauge field for running dilaton solutions is secondary. The fact that the AdS radii are different implies that the dual theories are different and each type of solutions requires its own holographic dictionary. 

Both types of solutions contain a number of arbitrary functions of time, as well as arbitrary constants. In the case of running dilaton solutions all arbitrary functions correspond to (pure gauge) sources of dual operators, while one of the arbitrary functions parameterizing the constant dilaton solutions corresponds to the one-point function of the scalar operator dual to the dilaton. Rather surprisingly, although this mode is identified with the one-point function of the scalar operator, it does not lead to a running dilaton, since it only enters in the solution for the metric and the gauge field. This mode is the only non trivial observable in the theory dual to constant dilaton solutions, but its significance and holographic interpretation have been mostly overlooked in the earlier literature, which is perhaps the reason behind the often made claim that AdS$_2$ holography with constant dilaton is trivial. 

In contrast, we find that this mode captures non trivial physics and leads to infinite families of qualitatively different solutions. In particular, negative values of this one-point function correspond to smooth horizonless geometries with two AdS$_2$ boundaries, while zero and positive values correspond respectively to extremal and non extremal black holes. The non extremal black holes with constant dilaton are distinct from those with running dilaton, but there is an RG flow that connects the extremal members of each family. This RG flow corresponds to the ``very-near-horizon region'' of (near) extremal black holes discussed in \cite{Strominger:1998yg}.      

Both types of 2D solutions can be uplifted to solutions of the STU model in 4D through the one-parameter family of Kaluza-Klein Ans\"atze given in \eqref{KK-relations}. The parameter $\l$ in the Ansatz corresponds to the rotation parameter of the resulting 4D solutions and can be viewed as a modulus. The uplift of either type of extremal solutions with zero $\l$ is a BPS solution, while for non zero $\l$ it is extremal non BPS. Running dilaton solutions uplift to general rotating subtracted geometries \cite{Cvetic:2011hp,Cvetic:2011dn}. As we mentioned above, these are generically non extremal asymptotically conformally AdS$_2\times S^2$ black holes that can be obtained through the near horizon decoupling limit of non extremal asymptotically flat black holes. Instead, the 4D solutions obtained by uplifting the constant dilaton solutions are asymptotically AdS$_2\times S^2$ and can be smooth geometries, extremal or non extremal black holes depending on the sign of the scalar one-point function. The BPS solution is qualitatively similar to the Near-Horizon-Extreme-Kerr (NHEK) geometry \cite{Bardeen:1999px}. In general, all 4D solutions resulting from the uplift of the constant dilaton solutions appear to be new asymptotically AdS$_2\times S^2$ solutions of the STU model.       

Our main result regarding the holographic dictionary is the identification of the dual operators and the Ward identities they satisfy. We consider Dirichlet boundary conditions on all fields for both running and constant dilaton solutions, but for constant dilaton solutions we also discuss Neumann boundary conditions on the gauge field, which uplift to Comp\`ere-Song-Strominger (CSS) boundary conditions in 3D \cite{Compere:2013bya}. Except for this case, the spectrum of operators in the dual theory consists of the stress tensor, the scalar operator dual to the dilaton and the current dual to the gauge field. For constant dilaton solutions with Neumann boundary conditions, however, there is no local current operator. Instead, there is a dynamical gauge field whose only gauge invariant observable is a non local `Polyakov line'. All these operators are necessary to consistently describe the physics of the dual theory.  

In particular, the scalar operator dual to the dilaton plays a major role for both running and constant dilaton solutions. For the running dilaton solutions it corresponds to a marginally relevant operator whose coupling drives the RG flow. Turning on an arbitrary source for this operator gives rise to a conformal anomaly that matches precisely the conformal anomaly of the 2D CFT that provides the UV completion. We find that the renormalized effective action for the sum of the stress tensor and the scalar operator, which equals the conformal anomaly through the trace Ward identity, can be expressed in terms of the Schwarzian derivative of a dynamical time on the boundary. With respect to the theory dual to constant dilaton solutions, this scalar operator has dimension 2 and is therefore irrelevant. However, the stress tensor vanishes identically in this theory and so the one-point function of this operator is the only non trivial observable, parameterizing the Coulomb branch of the theory. Moreover, it transforms anomalously under local conformal transformations, with a central charge related to that of the UV completion.    

While our holographic dictionary for running dilaton solutions mostly agrees with previous works, our analysis for the constant dilaton solutions differs considerably. This is due to the fact that we use different boundary counterterms than previous studies, which are dictated by the asymptotic behavior of the gauge field in the case of constant dilaton solutions. We explain at length in section \ref{dict} how general consistency arguments for the counterterms, that go beyond the requirement that the divergences of the on-shell action be removed, unambiguously lead to our counterterms. In particular, these boundary counterterms are crucial for AdS$_2$ holographic correlation functions to be consistent \cite{usfuture}. 

By uplifting the running and constant dilaton solutions to solutions of 3D gravity with a negative cosmological constant, we show that the former are obtained by a {\em spacelike} circle reduction of the general solution of 3D gravity, while the latter correspond to a {\em null} reduction \cite{Castro:2014ima}. This is partly why the boundary counterterms for running dilaton solutions can be obtained by Kaluza-Klein reduction from those of 3D gravity, but the counterterms for constant dilaton solutions cannot. Since the 2D solutions solve all the equations of motion following from the action \eqref{action}, their uplift to 3D automatically solves the Einstein equations, including the trace and divergence constraints on the holographic stress tensor. In the case of constant dilaton solutions this leads to a holographic stress tensor that contains an arbitrary function and precisely matches the stress tensor for CSS boundary conditions \cite{Compere:2013bya}. From the 2D point of view, this arbitrary function corresponds to the one-point function of the irrelevant scalar operator dual to the dilaton.

Turning to the symmetries preserved by each type of solutions, we find that running dilaton solutions always admit a single timelike Killing vector, recovering a well known result from the earlier literature. In combination with the electric charge, therefore, the classical symmetry algebra for such solutions is $u(1)\oplus u(1)$. The corresponding conserved charges are respectively the mass and the electric charge. In contrast, constant dilaton solutions admit an infinite set of isometries that generate a Witt algebra (classical Virasoro), but this is broken to 
the global $sl(2,\bb R)$ subalgebra by the anomalous transformation of the scalar operator. In addition, there is a global electric $u(1)$ in the case of Dirichlet boundary conditions on the gauge field, or a local $\Hat u(1)$ Kac-Moody algebra in the case of Neumann boundary conditions. The fact that the stress tensor is identically zero for constant dilaton solutions leads to the conclusion that all charges associated with the conformal symmetry vanish for both boundary conditions on the gauge field. Moreover, for Dirichlet boundary conditions, the charge corresponding to the global $u(1)$ symmetry is the non zero electric charge, while for Neumann boundary conditions there is no conserved charge associated with the $\Hat u(1)$ Kac-Moody symmetry. 

Finally, we consider the asymptotic symmetries preserved by the 2D solutions when uplifted to three dimensions. This allows for more general symmetry transformations that involve the Kaluza-Klein circle, and consequently for extended symmetry algebras, since the Killing vectors are now solutions of partial differential equations. Although such generalized isometries generate infinite dimensional algebras for both running and constant dilaton solutions, we find that the only non trivial conserved charges for running dilaton solutions are still the mass and the electric charge, which correspond respectively to the Hamiltonians along the time and Kaluza-Klein circle directions. However, the fact that constant dilaton solutions contain the arbitrary one-point function of the scalar operator allows for a non trivial realization of the infinite dimensional symmetry algebras. In particular, Dirichlet boundary conditions on the gauge field lead to non trivial conserved charges for one copy of the Virasoro algebra, as well as an electric $u(1)$. For Neumann boundary conditions on the gauge field, which correspond to CSS boundary conditions from the 3D point of view, we find non trivial charges for one copy of the Virasoro algebra, as well as a $\Hat u(1)$ Kac-Moody algebra, with the central charge and Kac-Moody level found in \cite{Compere:2013bya}, in agreement with \cite{Castro:2014ima}.

\paragraph{Organization} The rest of the paper is organized as follows. In section \ref{model} we present the most general solutions of the action \eqref{action}, with either running or constant dilaton, and discuss the corresponding vacuum and black hole solutions. Moreover, we discuss the uplift of any solution of the action \eqref{action} to a solution of the STU model \eqref{STU-Action-Reduced} in four dimensions through the Kaluza-Klein Ansatz \eqref{KK-relations}. The holographic dictionary for both running and constant dilaton solutions is constructed in section \ref{dict}, while is section \ref{CFT2} we uplift the general 2D solutions to solutions of 3D Einstein-Hilbert gravity with negative cosmological constant. This allows us to find the precise map between the 2D and 3D holographic dictionaries. The asymptotic symmetries and the corresponding conserved charges for both classes of solutions are computed in section \ref{algebra}. In section \ref{3Dsymmetries} we consider more general asymptotic transformations that involve the circle direction from three dimensions, which leads to extended symmetries and non trivial infinite dimensional algebras in the case of constant dilaton solutions. Finally, in appendix \ref{appendix} we summarize some essential results on subtracted geometries and present our Kaluza-Klein Ansatz for the reduction of the 4D STU model to two dimensions, while in appendix \ref{appendixB} we discuss the connection of our results to those of \cite{Castro:2014ima} and \cite{Hartman:2008dq}.

\section{The general solution of 2D Einstein-Maxwell-Dilaton gravity}
\label{model}
\setcounter{equation}{0}

In this section we obtain the most general solutions of the 2D action \eqref{action}, including the general solution with constant dilaton found earlier in \cite{Castro:2008ms}, and discuss their connection with 4D asymptotically conformally AdS$_2\times S^2$  black holes. The equations of motion following from the action \eqref{action} take the form 
\bsub
\label{eoms}
\bal
&0= R[g]+\frac{2}{L^2}-\frac34 e^{-2\j}F_{ab}F^{ab},\\
&0=\(\nabla_a\nabla_b-g_{ab}\square\)e^{-\j}+\frac12e^{-3\j}\Big(F_{ac}F_{b}{}^c-\frac14g_{ab}F_{cd}F^{cd}\Big)+\frac{1}{L^2}g_{ab}e^{-\j},\\
&0= \nabla_a(e^{-3\j}F^{ab}).
\eal
\esub
Without loss of generality, we choose to work in the Fefferman-Graham gauge\footnote{From the radial Hamiltonian formulation of the bulk dynamics in section \ref{dict} follows that this form of the metric can always be reached locally by a bulk diffeomorphism, and so it corresponds to a choice of gauge. The theorem of Fefferman and Graham ensures that this choice of gauge is always possible in the vicinity of the conformal boundary, but it may break down in the interior. However there is no loss of generality in seeking the general solution in the gauge \eqref{FG}. Another gauge choice often used is conformal gauge. The unique advantage of the gauge \eqref{FG} is that it allows us to identify the general boundary data corresponding to local sources in the dual theory.} 
\be
\label{FG}
ds^2=du^2+\g_{tt}(u,t)dt^2,
\ee
so that the equations of motion become
\bsub
\label{eoms-FG}
\bal
&\(\pa_u^2-L^{-2}\)e^{-\j}+Q^2e^{3\j}=0,\label{eoms-2}\\
&\(\pa_u^2-L^{-2}-3Q^2e^{4\j}\)\sqrt{-\g}=0,\label{eoms-1}\\
&\(\square_t+K\pa_u-L^{-2}\)e^{-\j}+Q^2e^{3\j}=0,\label{eoms-ham}\\
&\pa_u\(\pa_t e^{-\j}/\sqrt{-\g}\)=0\label{eoms-mom},\\
&\pa_uQ=\pa_tQ=0\label{eoms-gauge},
\eal
\esub
where $\g_{tt}=-(\sqrt{-\g})^2$, $K=\pa_u\log\sqrt{-\g}$ is the extrinsic curvature of $\g_{tt}$, $\square_t$ is the covariant Laplacian with respect to $\g_{tt}$, and $Q=\frac12\sqrt{-\g} e^{-3\j}F^{ut}$. 

The general solution of this system of equations can be obtained analytically. The equations of motion imply that $Q$ is a constant and hence equation 
\eqref{eoms-2} is a decoupled second order non linear equation for the dilaton $\j$, known as Yermakov’s equation (see \cite{Polyanin}, 2.9.1-1). The general solution of this equation consists of a continuous family of solutions with a non trivial dilaton profile, as well as an isolated solution with constant dilaton. These two solutions correspond to different Fefferman-Graham asymptotic expansions and hence define different holographic duals. We will therefore analyze the holographic dictionary for each of these asymptotic solutions separately.

\subsection{General solution with running dilaton}

The general solution of equations \eqref{eoms-FG} with running dilaton takes the form
\bsub
\label{gen-sol}
\bal
e^{-\j}&=\b(t)e^{u/L}\sqrt{\(1+\frac{m-\b'^2(t)/\a^2(t)}{4\b^2(t)}L^2e^{-2u/L}\)^2-\frac{Q^2L^2}{4\b^4(t)}e^{-4u/L}},\\
\sqrt{-\g}&=\frac{\a(t)}{\b'(t)}\pa_t e^{-\j},\\
A_t&=\m(t)+\frac{\a(t)}{2\b'(t)}\pa_t\log\(\frac{4L^{-2}e^{2u/L}\b^2(t)+m-\b'^2(t)/\a^2(t)-2Q/L}{4L^{-2}e^{2u/L}\b^2(t)+m-\b'^2(t)/\a^2(t)+2Q/L}\),
\eal
\esub
where $\a(t)$, $\b(t)$ and $\m(t)$ are arbitrary functions of time, while $m$ and $Q$ are arbitrary constants. Moreover, the primes ${}'$ denote a derivative with respect to $t$. The subclass of solutions \eqref{gen-sol} with $Q=0$ correspond to the general solution of the Jackiw–Teitelboim model \cite{Jackiw:1982hg,Teitelboim:1983fg}. Note that equations \eqref{eoms-FG} are symmetric under $u\to-u$ and so replacing $u$ with $-u$ in \eqref{gen-sol} gives another, equivalent solution. Although the expressions for the metric and gauge field may at first sight look ill defined when the arbitrary functions of time are set to constants, they do in fact admit a smooth limit, which is given in \eqref{static-sol} below. 

The leading asymptotic behavior of the solution \eqref{gen-sol} as $u\to+\infty$ is 
\be
\label{sources-I}
\g_{tt}= -\a^2(t)e^{2u/L}+\co(1),\quad e^{-\j}\sim \b(t)e^{u/L}+\co(e^{-u/L}),\quad A_t=\m(t)+\co(e^{-2u/L}),  
\ee 
and so, as we will discuss in more detail in section \ref{dict}, the arbitrary functions $\a(t)$, $\b(t)$ and $\m(t)$ should be identified with the sources of the corresponding dual operators. A special property of this particular class of solutions of 2D dilaton gravity is that all sources are pure gauge. In particular, the arbitrary functions $\a(t)$, $\b(t)$ and $\m(t)$ can be eliminated by means of a bulk diffeomorphism and a U(1) gauge transformation. As a consequence, the holographic one-point functions that we will obtain from this class of solutions are local functions of the sources and the Ward identities are explicitly satisfied. Nevertheless, these sources are important in order to describe the holographic dictionary and the mechanism responsible for the breaking of conformal invariance in the dual one-dimensional theory, and it is this type of `pure gauge dynamics' that has attracted attention recently \cite{Almheiri:2014cka,Maldacena:2016upp,Jensen:2016pah,Engelsoy:2016xyb}. However, the parameters $m$ and $Q$ describe genuine dynamics. 

If the arbitrary functions $\a(t)$, $\b(t)$ and $\m(t)$ are set to some constant values, respectively $\a_o$, $\b_o$ and $\m_o$, the solution \eqref{gen-sol} becomes \cite{Strominger:1998yg}
\bsub
\label{static-sol}
\bal
e^{-\j}&=\b_o e^{u/L}\sqrt{\(1+\frac{mL^2}{4\b^2_o}e^{-2u/L}\)^2-\frac{L^2Q^2}{4\b_o^4}e^{-4u/L}},\\
\sqrt{-\g}&=\frac{\a_oL}{\b_o}\;\pa_u e^{-\j},\\
A_t&=\m_o+\frac{\a_o}{\b_o}\; LQ e^{2\j}.
\eal
\esub
For generic $m>0$ and $|Q|< mL/2$, this is a non extremal asymptotically AdS$_2$ black hole, which becomes extremal when $Q=\pm mL/2$. The minimum value of the radial coordinate $u$ corresponds to the outer horizon, where the induced metric $\g_{tt}$ vanishes, and is given by 
\be
e^{2u_+/L}=\frac{L}{4\b^2_o}\sqrt{m^2L^2-4Q^2}.
\ee
In the extremal case $u_+=-\infty$ and so $u$ goes from $-\infty$ to $+\infty$. The value of the dilaton on the outer horizon is
\be
e^{-\j(u_+)}=\frac{L^{1/2}}{2}\(\sqrt{mL+2Q}+\sqrt{mL-2Q}\).
\ee

The Hawking temperature can be computed as usual by requiring that the Euclidean section does not have a conical defect. The result is   
\be\label{T}
T=\frac{\a_o\b_o}{\p L^{1/2}}\frac{\sqrt{m^2L^2-4Q^2}}{\sqrt{mL+2Q}+\sqrt{mL-2Q}},
\ee
which indeed vanishes when $m=2Q/L$. However, in two dimensions the entropy is not given by the area law, but can be computed for example using Wald's formula \cite{Wald:1993nt,Iyer:1994ys}. For black holes of generic 2D dilaton gravity models ones finds that the entropy is given by the value of the dilaton on the horizon \cite{Myers:1994sg,Gegenberg:1994pv,Cadoni:1999ja}: 
\be\label{S}
S=\frac{2\p}{\k_2^2}e^{-\j(u_+)}.
\ee

\subsection{General solution with constant dilaton}

A distinct class of solutions of the field equations \eqref{eoms-FG} is \cite{Castro:2008ms,Castro:2014ima,Jensen:2016pah}
\bsub\label{gen-sol-2}
\bal
e^{-2\j}&=LQ,\\
\sqrt{-\g}&=\wt\a(t)e^{u/\wt L}+\frac{\wt\b(t)}{\sqrt{LQ}}e^{-u/\wt L} ,\\
A_t&=\wt\m(t)-\frac{1}{\sqrt{LQ}}\Big(\wt\a(t)e^{u/\wt L}-\frac{\wt\b(t)}{\sqrt{LQ}}e^{-u/\wt L}\Big),
\eal
\esub
where $\wt\a(t)$, $\wt\b(t)$ and $\wt\m(t)$ are arbitrary functions, $Q>0$ is an arbitrary constant, and $\wt L=L/2$. Note that, contrary to the running dilaton solutions, this class of solutions involves the gauge field in an essential way, since the electric charge must be non zero.

A number of distinctive properties of this class of solutions should be emphasized. Although both solutions \eqref{gen-sol} and \eqref{gen-sol-2} are asymptotically locally AdS$_2$, the AdS radius, $\wt L$, of \eqref{gen-sol-2} is half that of \eqref{gen-sol}. Since the degrees of freedom of the holographic dual is generically proportional to a positive power of the AdS radius (e.g. $N^2\sim L^3$ in $\cn=4$ $SU(N)$ super Yang-Mills in four dimensions, or $c\sim L$ in two dimensional CFTs with a gravity dual \cite{Brown:1986nw}), this suggests that there may exist an interpolating flow between \eqref{gen-sol} in the UV and \eqref{gen-sol-2} in the IR, corresponding to a renormalization group (RG) flow between two different theories. As we show in the next subsection, there is indeed an interpolating flow between the extremal elements of each family of solutions, but not for the non extremal ones.

Another property of the constant dilaton solutions \eqref{gen-sol-2} that is worth pointing out is that the asymptotic form of the gauge field $A_t$ is rather unlike that of gauge fields in AdS$_{d+1}$ with $d>2$, or the solution \eqref{gen-sol} in the case of running dilaton, since the mode that asymptotically dominates is not $\wt\m(t)$, but rather $\wt\a(t)$. This phenomenon occurs generically for antisymmetric $p$-form fields in AdS$_{d+1}$ with $p\geq d/2$ \cite{thermo2} and it is the source of some confusion in the literature regarding the correct holographic dictionary and holographic renormalization in these cases, particularly in the study of one-form gauge fields in AdS$_2$ and AdS$_3$.    
We will discuss the holographic dictionary in detail in section \ref{dict}.

Finally, the asymptotic form of the solution \eqref{gen-sol-2} suggests that the arbitrary functions $\wt\a(t)$ and $\wt\m(t)$ correspond to the sources of the dual stress tensor and U(1) current,\footnote{Although $\wt\m(t)$ is not the leading mode in the asymptotic expansion of the gauge field, it is the only mode that defines a local operator in the dual theory. We will return to this point in section \ref{dict}.} but the role of the arbitrary function $\wt\b(t)$ is less obvious. As we will see in section \ref{dict}, it corresponds to the one-point function of the scalar operator dual to the dilaton $\j$, which is an irrelevant operator of dimension 2 relative to the theory dual to constant dilaton solutions \eqref{gen-sol-2}. This mode has been discussed before, e.g. in \cite{Castro:2008ms,Balasubramanian:2009bg}, but its holographic interpretation was different due to the use of different boundary counterterms. We will address to this point in detail in section \ref{dict}. 

Our analysis in sections \ref{dict} and \ref{algebra} implies that $\wt\b$ parameterizes degenerate states in the dual one dimensional theory, and the 
identification of this mode with the one-point function of a scalar operator  suggests that it describes excitations of the Coulomb branch of the dual theory \cite{Strominger:1998yg}. Moreover, it seems plausible that it is related to AdS$_2$ fragmentation \cite{Maldacena:1998uz}. In particular, when $\wt\b>0$, the geometry \eqref{gen-sol-2} is smooth and another boundary opens up at $u\to -\infty$. As we will see in the next subsection, starting with the extremal solution with running dilaton, one can reach a solution of the form \eqref{gen-sol-2} in the IR with either $\wt\b$ or $\wt\a$ set to zero, but not both non zero. The extremal solution with running dilaton, therefore, zooms in on one of the AdS$_2$ throats.\footnote{The mode $\wt\b$ is normalizable with respect to the boundary at $u=+\infty$, but non normalizable with respect to that at $u=-\infty$ \cite{Balasubramanian:2009bg}. The opposite holds for $\wt\a$. However, our interpretation of this mode refers to an observer in the dual theory living at $u=+\infty$. From the point of view of this observer all $\wt\b>0$ lead to smooth bulk geometries and so $\wt\b$ can be arbitrary, perceived as massless excitations of the theory living at $u=+\infty$. Conversely, $\wt\a$ parameterizes massless excitations of the theory living at $u=-\infty$. Hence, although there are no excitations of AdS$_2$ that are normalizable at both boundaries, observers on each of the boundaries do see massless excitations, parameterized by the one-point function of the dimension 2 scalar operator dual to the dilaton.}

For $\wt\b=0$, \eqref{gen-sol-2} describes an extremal black hole, while $\wt\b<0$ corresponds to a non extremal black hole. An important feature of these black holes is that their thermodynamic properties are {\em not} the same as those of the BTZ black hole one obtains by uplifting them to three dimensions. This is in stark contrast to the behavior of running dilaton black holes, which have the same thermodynamic properties as their 3D uplifts. The reason behind this property of constant dilaton black holes is that, as we will see in section \ref{CFT2}, they are obtained by a {\em null} reduction from three dimensions \cite{Castro:2014ima}, while running dilaton solutions are obtained by a spacelike reduction. Therefore, the difference in the thermodynamics of constant dilaton black holes arises from the fact that what is time in 2D is a null coordinate in 3D.   

Setting $\wt\a$ and $\wt\b$ to constants, the thermodynamic observables of the constant dilaton black holes, computed directly in 2D, are    
\be\label{2D-thermo}
T\sbtx{2D}=\frac{\sqrt{-\wt\a_o\wt\b_o}}{\p \wt L(LQ)^{1/4}}\;,\qquad S\sbtx{2D}=\frac{2\p}{\k_2^2}\sqrt{LQ}\;,\qquad M\sbtx{2D}=0\;.
\ee
These agree with the expressions given e.g. in \cite{Castro:2008ms,Jensen:2016pah}, despite the fact that we use different boundary counterterms. Uplifting these black holes to 3D using the results of section \ref{CFT2} and computing the thermodynamic observables of the resulting BTZ black hole we find instead
\be\label{3D-thermo}
T\sbtx{3D}=\frac{(LQ)^{1/4}\sqrt{-\wt\a_o\wt\b_o}}{\p \wt L\Big(\sqrt{LQ}+\sqrt{\frac{-2\wt\b_o}{LQ}}\Big)},\quad S\sbtx{3D}=\frac{2\p}{\k_2^2}\Big(\sqrt{LQ}+\sqrt{\frac{-2\wt\b_o}{LQ}}\;\Big),\quad M\sbtx{3D}=\frac{1}{4\k^2_2\wt L}\Big(LQ-\frac{2\wt\b_o}{LQ}\Big).
\ee
We would like to view these as the correct expressions describing the thermodynamics of the constant dilaton black holes and they may be an indication that the Kaluza-Klein circle cannot be really ignored in the case of constant dilaton solutions. We will find further evidence to this effect in sections \ref{algebra} and \ref{3Dsymmetries}.

\subsection{Extremal solution as an interpolating RG flow}

Setting $m-\b'^2/\a^2=2Q/L>0$ and $\m=-\a/\b$ in \eqref{gen-sol} the solution becomes 
\be\label{RG-flow}
e^{-\j}=\sqrt{LQ+\b^2(t) e^{2u/L}},\qquad
\sqrt{-\g}=\frac{\a(t) \b(t)e^{2u/L}}{\sqrt{LQ+\b^2(t)e^{2u/L}}},\qquad
A_t=-\frac{\a(t) \b(t)e^{2u/L}}{LQ+\b^2(t)e^{2u/L}}.
\ee
The asymptotic form of this as $u\to+\infty$ is still given by \eqref{sources-I}, but for $u\to -\infty$ it behaves as
\bsub
\label{RG-flow-exp}
\bal
e^{-\j}&=\sqrt{LQ}+\frac{\b^2}{2\sqrt{LQ}}e^{2u/L}+\co(e^{4u/L}),\\
\sqrt{-\g}&=\frac{\a\b}{\sqrt{LQ}}\; e^{2u/L}\Big(1-\frac{\b^2}{2LQ}e^{2u/L}+\co(e^{4u/L})\Big),\\
A_t&=-\frac{\a\b}{LQ} e^{2u/L}\Big(1-\frac{\b^2}{LQ}e^{2u/L}+\co(e^{4u/L})\Big).
\eal
\esub
In particular, the limit $\b\to 0$ keeping $\a\b$ fixed results in an exact solution of the form \eqref{gen-sol-2} with $\wt\a=\a\b/\sqrt{LQ}$ and $u\to -u$. Notice that this limit sets $m=2Q/L$ and $\m\to-\infty$, and corresponds to the 
``very-near-horizon region'' discussed in \cite{Strominger:1998yg}. The solution \eqref{RG-flow} describes an RG flow between the theory dual to running dilaton asymptotics and that dual to constant dilaton asymptotics. We will derive the holographic dictionary for both theories in section \ref{dict}. 

\subsection{Vacuum solutions}

Provided the electric charge $Q$ is zero, the equations of motion \eqref{eoms-FG} admit exact AdS$_2$ solutions with a non trivial scalar profile, namely  
\bsub
\label{instantons}
\bal
e^{-\j}&=(\b_o+\b_1 t+\b_2 t^2)e^{u/L}-\b_2 e^{-u/L},\label{inst-1}\\
\sqrt{-\g}&=e^{u/L},\\
A_t&=\m(t),
\eal
\esub
where $\b_o$, $\b_1$ and $\b_2$ are constants. Such solutions are a characteristic property of conformally coupled scalars and exist even in higher dimensions  \cite{deHaro:2006ymc,Papadimitriou:2007sj}. The Euclidean version of these solutions (obtained by flipping the sign of the last term in \eqref{inst-1}) was the focus of the analysis of \cite{Maldacena:2016upp}, where hyperbolic space was cut off by prescribing a boundary condition on the scalar.  

In the present context, \eqref{instantons} corresponds to the vacuum of the theory dual to the running dilaton solutions, with the scalar field allowed to take arbitrarily large values. This results in the breaking of the AdS$_2$ isometry group from $SL(2,\bb R)$ to $U(1)$ \cite{Maldacena:2016upp,Engelsoy:2016xyb}, in direct analogy with D$p$ branes for $p\neq 3$, where the running of the dilaton breaks the AdS$_{p+2}$ isometry group to $ISO(1,p)$ \cite{Itzhaki:1998dd,Kanitscheider:2008kd}.

For the theory dual to the constant dilaton solutions \eqref{gen-sol-2}, the vacua are parameterized by strictly positive constant values of $\wt \b$, which corresponds to the Coulomb branch of the theory. The origin of the Coulomb branch is at $\wt\b_o=\sqrt{LQ}\;\wt\a_o$, which is global AdS$_2$. Depending on the boundary conditions on the gauge field, the symmetry preserved at the origin of the Coulomb branch is either a global $u(1)$ and a classical Virasoro algebra (Witt algebra), or a classical Kac-Moody $u(1)$ and a classical Virasoro. However, both local symmetries are broken to their global subalgebras due to anomalies. We will return to the discussion of the asymptotic symmetries and of the corresponding conserved charges in sections \ref{algebra} and \ref{3Dsymmetries}.

\subsection{Uplift to four dimensions}

Before we turn to the holographic dictionary for the two classes of solutions of the 2D model \eqref{action}, it is instructive to spell out the connection of such solutions with 4D black holes. Here we will use notation that is introduced in appendix \ref{appendix} and \cite{An:2016fzu}.

Both the running dilaton solutions \eqref{gen-sol} and the constant dilaton solutions \eqref{gen-sol-2} can be uplifted to solutions of the STU model \eqref{STU-Action-Reduced} in four dimensions, as is indicated schematically in Fig. \ref{fig1}. The explicit Kaluza-Klein Ansatz is given in \eqref{KK-relations} and contains a free parameter $\l$, which is related to the angular velocity of the resulting 4D solutions. In particular, a given 2D solution can be uplifted to a family of 4D solutions with different angular velocities. 

The running dilaton solutions uplift to so called `subtracted geometries' \cite{Cvetic:2011hp,Cvetic:2011dn,Cvetic:2012tr,Virmani:2012kw,Baggio:2012db,Cvetic:2013vqi,Cvetic:2014sxa}. As was shown in \cite{Baggio:2012db} for the static solutions, the subtracted geometries can be obtained (besides the original subtraction procedure \cite{Cvetic:2011hp,Cvetic:2011dn}, scaling limits \cite{Cvetic:2012tr}, and Harrison transformations \cite{Virmani:2012kw,Cvetic:2013vqi}) through a {\em decoupling} limit of generic multi-charge non extremal asymptotically flat black holes of the STU model in four dimensions \cite{Cvetic:1995kv,Cvetic:1996kv,Chong:2004na,Chow:2013tia,Chow:2014cca}. In particular, the subtracted geometries can be obtained by setting to zero certain integration constants in the harmonic functions of asymptotically flat black holes, much in the same way that an asymptotically conformally AdS$_{p+2}\times S^{8-p}$ solution can be obtained by setting to zero the integration constant in the harmonic function of D$p$ branes. The resulting static or stationary subtracted geometries are asymptotically conformally AdS$_2\times S^2$ (equivalently asymptotically conical \cite{Cvetic:2012tr}), generically {\em non extremal}, black holes of the STU model and take the form \eqref{subtracted-simple}. The relation between the radial coordinates in \eqref{gen-sol} and \eqref{subtracted-simple} is   
\be
4r=e^{2(u-u_o)/L}+2(r_++r_-)+(r_+-r_-)^2e^{-2(u-u_o)/L},\quad e^{-u_o/L}=\frac{4\ell}{L},
\ee
while the various constants that parameterize the solutions are mapped as
\be
L=2B,\qquad \k_2^2=\k_4^2/\p L^2,
\ee  
\be\label{source-map}
\a_o=k,\quad \b_o=\(\ell/B\)^3,\quad \m_o=0,
\ee
and
\be\label{vev-map}
mL^2=\(2\ell/L\)^4\(r_++r_-\),\quad LQ=\(2\ell/L\)^4\sqrt{r_+r_-}\;.
\ee
In particular, the integration constants $k$ and $\ell$ that were introduced in \cite{An:2016fzu} in order to formulate the variational problem for the stationary solutions correspond to the sources of the operators dual to the 2D metric and dilaton, respectively. Moreover, since $\l$ enters in the azimuthal angle of the internal $S^2$ in the reduction Ansatz \eqref{KK-relations}, the combination $kL\o$ must be an integer for the internal $S^2$ to be free of conical singularities \cite{An:2016fzu}. Finally, inserting the relations \eqref{source-map} and \eqref{vev-map} in the expressions \eqref{T} and \eqref{S} for the temperature and the entropy of the 2D black hole gives respectively the temperature and entropy of the 4D black hole.

\begin{figure}
	\begin{tabular}{ccc}
		\scalebox{0.5}{\hskip-0.0in \includegraphics{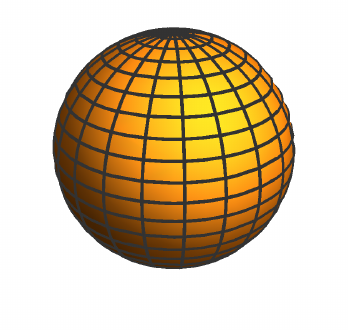}} &\hskip0.5in
		\scalebox{0.5}{\hskip-0.0in 
			\includegraphics{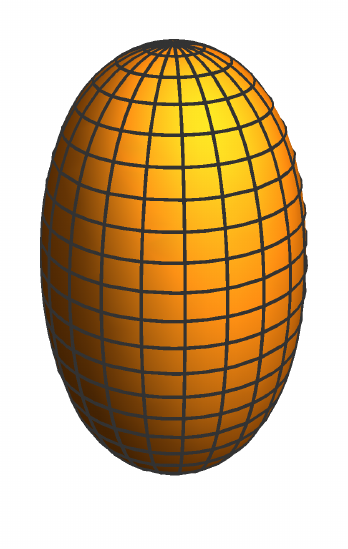}}&\hskip0.5in  
		
		\scalebox{0.5}{\hskip-0.0in \includegraphics{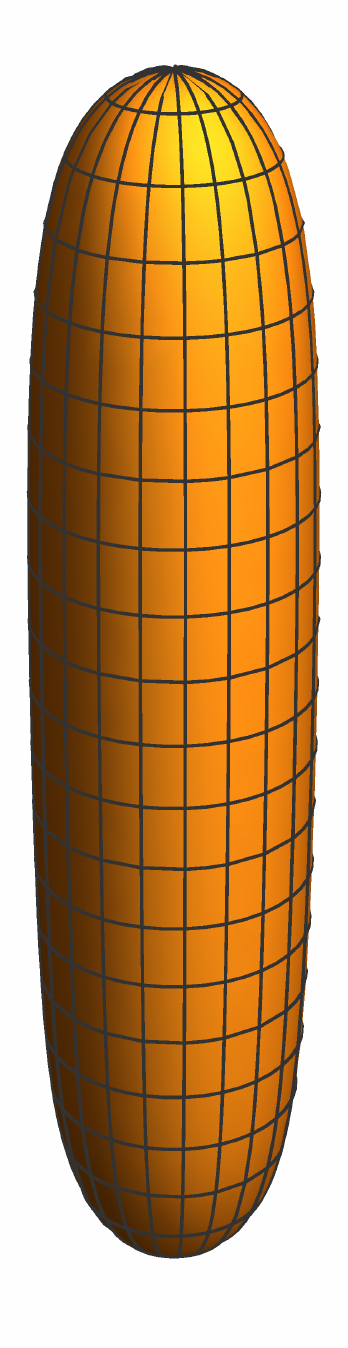}}\\
		(a) &\hskip0.5in  (b) &\hskip0.5in (c)
	\end{tabular}
	\centering
	\caption{The compact part of the solutions \eqref{classII-uplift} for increasing values of the ratio $\l B/\sqrt{LQ}$: 0 for plot (a), 1 for plot (b), and 3 for plot (c). Solutions with $\l=0$ are static, but when $\l\neq 0$ the solutions acquire a non zero angular velocity, which goes to infinity at the AdS$_2$ boundary. }
	\label{fig2}
\end{figure}

The Kaluza-Klein Ansatz \eqref{KK-relations} also allows us to uplift the constant dilaton solutions \eqref{gen-sol-2}, resulting in novel 4D solutions, which to our knowledge are new. These solutions are generically asymptotically AdS$_2$ times a deformed $S^2$, instead of conformally AdS$_2\times S^2$ as was the case for the subtracted geometries that are obtained by uplifting the running dilaton solutions. Moreover, the AdS$_2$ radius for these solutions is half that of the AdS$_2$ appearing in the subtracted geometries. The explicit form of the 4D uplift of constant dilaton solutions is  
\bsub\label{classII-uplift}
\bal
e^{-2\h}&=LQ+\l^2B^2\sin^2\th,\quad
\c=\l B\cos\th, \quad
A+\c A^0=B\cos\th d\f, \\
e^{-2\h}A^0&=-\sqrt{LQ}\Big(\wt\a(t)e^{u/B}-\frac{\wt\b(t)}{\sqrt{LQ}}e^{-u/B}-\sqrt{LQ}\wt\m_o\Big)dt+\l B^2\sin^2\th d\f,\\
e^\h ds_4^2&=du^2-\Big(\wt\a(t)e^{u/B}+\frac{\wt\b(t)}{\sqrt{LQ}}e^{-u/B}\Big)^2dt^2\\
&\hskip-0.5cm+B^2\Big(d\th^2+\frac{LQ\sin^2\th}{LQ+\l^2B^2\sin^2\th}\Big(d\f+\frac{\l}{\sqrt{LQ}} \Big(\wt\a(t)e^{u/B}-\frac{\wt\b(t)}{\sqrt{LQ}}e^{-u/B}-\sqrt{LQ}\;\wt\m(t)\Big)dt\Big)^2\Big).\NO
\eal
\esub
For $\l=0$ and $\wt\b=0$ this is the Robinson-Bertotti geometry on AdS$_2\times S^2$ and corresponds to an extremal BPS black hole solution in four dimensions. As in 2D, this extremal solution arises as the far IR limit of the corresponding extremal subtracted geometry. For non zero $\l$ and $\wt\b=0$ this solution is still extremal but not BPS. The internal $S^2$ gets deformed as shown in Fig.~\ref{fig2} and it acquires non zero angular velocity, which becomes infinite near the boundary of AdS$_2$. This solution is analogous to -- but distinct from -- the NHEK geometry \cite{Bardeen:1999px} in pure gravity. 

Finally, for $\wt\b<0$ these correspond to non extremal black holes, that are distinct from the subtracted geometries, while for $\wt\b>0$ they are smooth horizonless geometries, which from the 2D perspective might be interpreted as excitations of the Coulomb branch of the dual CFT$_1$. It would be interesting to see if these solutions can be generalized to asymptotically flat horizonless geometries by restoring the integration constants in the harmonic functions, thus providing microstates for the corresponding asymptotically flat black holes.   

\section{Radial Hamiltonian formulation}
\label{dict}
\setcounter{equation}{0}

In this section we derive the holographic dictionary for both running and constant dilaton solutions of the EMD theory \eqref{action}. In particular, using the radial Hamiltonian formulation of the dynamics, we identify the sources and the dual operators, determine the boundary counterterms required for each type of solutions, and evaluate the renormalized  one-point functions and on-shell action. The general solutions \eqref{gen-sol} and \eqref{gen-sol-2} allow us to obtain these quantities exactly as functions of arbitrary sources, and hence any higher $n$-point function can be directly obtained by further differentiation with respect to the sources.  

Inserting the radial decomposition 
\be\label{ADM-metric}
ds^2=(N^2+N_tN^t)du^2+2N_tdu dt+\g_{tt}dt^2,
\ee
of the metric in the action \eqref{action} we obtain the radial Lagrangian 
\bal\label{lagrangian}
\cl=\frac{1}{2\k^2_2}\int \tx dt\sqrt{-\g}N\Big(-\frac{2}{N}K\(\dot\j-N^t\pa_t\j\)
-\frac{1}{2N^2}e^{-2\j}F_{ut}F_{u}{}^t
+\frac{2}{L^2}-2\square_t\Big)e^{-\j},
\eal
where $K=\g^{tt}K_{tt}$ and the extrinsic curvature $K_{tt}$ is given by
\be
K_{tt}=\frac{1}{2N}\left(\dot\g_{tt}-2D_tN_t\right),
\ee
with the dot denoting a derivative with respect to the radial coordinate $u$, and $D_t$ standing for the covariant derivative with respect to the induced metric $\g_{tt}$.

The canonical momenta obtained from \eqref{lagrangian} are 
\bsub
\label{momenta}
\bal
\p^{tt}&=\frac{\d \cl}{\d\dot\g_{tt}}=-\frac{1}{2\k^2_2}\sqrt{-\g}e^{-\j}
\frac{1}{N}\g^{tt}\left(\dot\j-N^t\pa_t\j\right),\\
\p^{t}&=\frac{\d \cl}{\d\dot A_{t}}=-\frac{1}{2\k^2_2}\sqrt{-\g}e^{-3\j}
\frac{1}{N}\g^{tt}F_{ut},\\
\p_\j&=\frac{\d \cl}{\d\dot\j}=-\frac{1}{\k^2_2}\sqrt{-\g}e^{-\j}K,
\eal
\esub
while those conjugate to $N$, $N_t$ and $A_u$ vanish identically. These fields are therefore non dynamical Lagrange multipliers imposing three first class constrains. In particular, the Legendre transform of the Lagrangian \eqref{lagrangian} gives the Hamiltonian 
\be
H=\int \tx dt\left(\dot\g_{tt}\p^{tt}+\dot A_t\p^t+\dot\j\p_\j\right)-\cl
=\int \tx dt\left(N\ch+N_t\ch^t+A_u\cf\right),
\ee
where 
\bsub
\label{constraints}
\bal	
\ch&=-\frac{\k^2_2}{\sqrt{-\g}}e^{\j}\left(2\p\p_\j+e^{2\j}\p^t\p_t\right)-\frac{\sqrt{-\g}}{\k^2_2}\(L^{-2}-\square_t\)e^{-\j},\\
\ch^t&=-2D_t\p^{tt}+\p_\j\pa^t\j,\\
\cf&=-D_t\p^t.
\eal
\esub	
Hence, the vanishing of the canonical momenta conjugate to $N$, $N_t$ and $A_u$  leads to the constraints 
\be\label{constraints0}
\ch=\ch^t=\cf=0,
\ee
which are identified respectively with the equations \eqref{eoms-ham}, \eqref{eoms-mom} and the second equation in \eqref{eoms-gauge}. 

In the Fefferman-Graham gauge  
\be\label{gauge-fixing}
N=1,\quad N_t=0,\quad A_u=0,
\ee
Hamilton's equations take the form
\bsub
\bal
\dot\g_{tt}&=\frac{\d H}{\d\p^{tt}}=-\frac{2\k^2_2}{\sqrt{-\g}}e^{\j}\p_\j\g_{tt},\\
\dot A_t&=\frac{\d H}{\d\p^{t}}=-\frac{2\k_2^2}{\sqrt{-\g}}e^{3\j}\p_t,\\
\dot\j&=\frac{\d H}{\d\p_\j}=-\frac{2\k_2^2}{\sqrt{-\g}}e^{\j}\p,\\\NO\\
\dot{\p}^{tt} &=-\frac{\d H}{\d\g_{tt}}=\frac{\k^2_2}{\sqrt{-\g}}e^{\j}\Big(\p^{tt}\p_\j
+\frac12e^{2\j}\p^t\p^t\Big)+\frac{\sqrt{-\g}}{2\k^2_2}\g^{tt}e^{-\j}L^{-2},\\
\dot{\p}^t &=-\frac{\d H}{\d A_t}=0,\\
\dot{\p}_\j &=-\frac{\d H}{\d\j}=\frac{\k^2_2}{\sqrt{-\g}}e^{\j}\left(2\p\p_\j
+3e^{2\j}\p^t\p_t\right)-\frac{\sqrt{-\g}}{\k^2_2}e^{-\j}L^{-2},
\eal
\esub
which reproduce the remaining three equations in \eqref{eoms-FG}.  

As a final ingredient in the derivation of the holographic dictionary, recall that the canonical momenta \eqref{momenta} can alternatively be expressed as gradients of Hamilton's principal function $\cs$ as 
\be\label{HJ-momenta}
\p^{tt}=\frac{\d\cs}{\d\g_{tt}},\quad \p^{t}=\frac{\d\cs}{\d A_{t}},\quad \p_\j=\frac{\d\cs}{\d\j},
\ee
where $\cs[\g,\j,A]$ is a functional of the induced fields $\g_{tt}$, $A_t$ and $\j$ and their $t$-derivatives only and coincides with the on-shell action. 
Inserting these expressions for the canonical momenta in the constraints \eqref{constraints} gives the Hamilton-Jacobi equations for $\cs$, which can be used to derive the covariant counterterms.

\subsection{Holographic dictionary for running dilaton solutions}

The radial Hamiltonian formulation of the EMD theory \eqref{action} allows us to systematically construct the holographic dictionary for any admissible boundary conditions. Starting with the running dilaton boundary conditions, we first need to determine the covariant boundary counterterms that render the variational problem well posed and the on-shell action finite. 

\paragraph{Boundary counterterms} For the running dilaton solutions the boundary counterterms can be determined through standard holographic renormalization, in a number of different ways. In particular, one can adapt the analysis for non conformal branes in \cite{Kanitscheider:2008kd}, or solve the Hamilton-Jacobi equation for $\cs$ asymptotically using the recursive procedure developed in \cite{Papadimitriou:2011qb}. In this specific case, however, the boundary counterterms can also be obtained by Kaluza-Klein reduction of the well known boundary counterterms for 3D Einstein-Hilbert gravity \cite{Henningson:1998gx}. The result is
\be\label{ct-running}
S\sbtx{ct}=-\frac{1}{\k_2^2}\int \tx dt\sqrt{-\g}\; L^{-1}\(1-u_oL\square_t\)e^{-\j}.
\ee
Notice that there is a counterterm that depends explicitly on the radial cutoff $u_o$, indicating the presence of a conformal anomaly. This counterterm is inherited from the analogous counterterm for 3D gravity, which is proportional to the Euler density of the boundary metric and is a topological quantity. This is reflected in the fact that the counterterm in \eqref{ct-running} that depends explicitly on the radial cutoff is in fact a total derivative. Consequently, it does not contribute to the renormalization of the one-point functions, but it is necessary in order to evaluate the on-shell action on a finite time interval. Moreover, this term is required for the consistency between the trace of the stress tensor and the transformation of the renormalized on-shell action under boundary Weyl transformations.

\paragraph{Dual operators} Having determined the boundary counterterm we can now evaluate the exact renormalized one-point functions in the presence of sources which are given by the renormalized canonical momenta \cite{Papadimitriou:2004ap}, namely
\be\label{ops-I}
\ct=2\Hat\p^t_t,\quad \co_\j=-\Hat\p_\j,\quad \cj^t=-\Hat\p^t,
\ee
where, using \eqref{ct-running} and the solution \eqref{gen-sol}, we find that the renormalized canonical momenta are 
\bsub
\label{ren-momenta}
\bal
\Hat\p^t_t&=\frac{1}{2\k^2_2}\lim_{u\to\infty}e^{u/L}\(
\pa_u e^{-\j}-e^{-\j}L^{-1}\)=-\frac{L}{4\k_2^2}\(\frac{m}{\b}-\frac{\b'^2}{\b\a^2}\),\\
\Hat\p^{t}&=\lim_{u\to\infty}\frac{e^{u/L}}{\sqrt{-\g}}\p^{t}=-\frac{1}{\k^2_2}\frac{Q}{\a},\\
\Hat\p_\j&=-\frac{1}{\k^2_2}\lim_{u\to\infty}e^{u/L}e^{-\j}\(K-L^{-1}\)=-\frac{L}{2\k^2_2}\(\frac{m}{\b}-\frac{\b'^2}{\b\a^2}-2\frac{\b'\a'}{\a^3}+2\frac{\b''}{\a^2}\).
\eal
\esub
Hence, there are three independent local operators in the holographic dual to the theory with running dilaton boundary conditions, whose exact one-point functions as a function of the arbitrary sources $\a(t)$, $\b(t)$ and $\m(t)$ take the form 
\be
\label{1pt-fns}\boxed{
\ct=-\frac{L}{2\k_2^2}\(\frac{m}{\b}-\frac{\b'^2}{\b\a^2}\),\quad
\cj^t=\frac{1}{\k^2_2}\frac{Q}{\a},\quad
\co_\j=\frac{L}{2\k^2_2}\(\frac{m}{\b}-\frac{\b'^2}{\b\a^2}-2\frac{\b'\a'}{\a^3}+2\frac{\b''}{\a^2}\).}
\ee
Note that we have refrained from using the canonical notation $\<\cdot\>$ for one-point functions  in order to emphasize the fact that these are the one-point functions in the presence of arbitrary sources, which can be taken as the {\em definition} of the dual operators \cite{Osborn:1991gm}. Moreover, as we shall see, all three operators \eqref{ops-I} are required in order to obtain a consistent holographic dictionary. However,  notice that for constant $\b$ the operators $\ct$ and $\co_\j$ are equal up to a sign, which is a direct consequence of the trace Ward identity \eqref{trace-WI}. 

\paragraph{Ward identities} The operators \eqref{1pt-fns} satisfy a number of general identities, independently of the values of the parameters $m$ and $Q$. In particular, the momentum and gauge constraints, i.e. equation \eqref{eoms-mom} and the second equation in \eqref{eoms-gauge}, imply that 
\be\label{WIs-I}
\pa_t\ct-\co_\j\pa_t\log\b=0,\qquad \cd_t\cj^t=0,
\ee
where $\cd_t$ denotes the covariant derivative with respect to the boundary metric $-\a^2$. These are identified respectively with the Ward identities associated with time reparameterization invariance and $U(1)$ gauge invariance of the dual theory. Moreover, the expressions \eqref{1pt-fns} for the one-point functions lead to the trace Ward identity
\be\label{trace-WI}
\ct+\co_\j=\frac{L}{\k_2^2}\(\frac{\b''}{\a^2}-\frac{\b'\a'}{\a^3}\)=\frac{L}{\k_2^2\a}\pa_t\(\frac{\b'}{\a}\)\equiv \ca.
\ee
This identity implies that the scalar operator $\co_\j$ is marginally relevant with an exact beta function $\b_\j=-1$. Moreover, as anticipated, there is a conformal anomaly that is proportional to the counterterm that explicitly depends of the cutoff, since from \eqref{gen-sol} we see that $-\sqrt{-\g}\;\square_t e^{-\j}\sim \pa_t\(\b'/\a\)$, precisely matching the expression for the conformal anomaly $\ca$. Notice that the conformal anomaly is
proportional to the source $\b$ of the scalar operator dual to the dilaton $\j$, and hence, the dilaton plays a central role in AdS$_2$ holography and the breaking of the symmetries of the vacuum.

\paragraph{Generating functional}

The renormalized one-point functions \eqref{1pt-fns} can be expressed as 
\be\label{1pt-fn-var-I}
\ct=\frac{\d S\sbtx{ren}}{\d\a},\quad \co_\j=\frac{\b}{\a}\frac{\d S\sbtx{ren}}{\d\b},\quad \cj^t=-\frac{1}{\a}\frac{\d S\sbtx{ren}}{\d\m},
\ee
in terms of the renormalized on-shell action
\be\label{Sren}\boxed{
S\sbtx{ren}[\a,\b,\m]=-\frac{L}{2\k_2^2}\int \tx dt\(\frac{m\a}{\b}+\frac{\b'^2}{\b\a}+\frac{2\m Q}{L}\)+S\sbtx{global},}
\ee
which is identified with the generating function of connected correlation functions in the dual theory. Since \eqref{Sren} is obtained by functionally integrating the relations \eqref{1pt-fn-var-I}, it is determined only up to an integration `constant' $S\sbtx{global}$, which is independent of the local sources $\a(t)$, $\b(t)$ and $\m(t)$. As indicated, $S\sbtx{global}$ captures global properties of the theory and can be determined by evaluating explicitly the renormalized on-shell action.\footnote{With the counterterm \eqref{ct-running} the expression for the renormalized on-shell action is
	\be\hskip-0.2cm
	S\sbtx{ren}=\frac{1}{2\k_2^2}\int \tx dt\(\lim_{u_o\to\infty}\left.\Big[\pa_u(\sqrt{-\g}\; e^{-\j})-\frac{2}{L}\sqrt{-\g}\(1-u_oL\square_t\)e^{-\j}\Big]\right|_{u_o} +\left.\(e^{-\j}\pa_u\sqrt{-\g}-\sqrt{-\g}\;\pa_u e^{-\j}\)\right|_{u_+}\),
	\ee 
where $u_+$ is the minimum value of the radial coordinate and $u_o$ is the radial cutoff. In order to compute the global part $S\sbtx{global}$ in \eqref{Sren}, both $u_+$ and $u_o$ need to be expressed in terms of a suitable `uniformizing' Liouville field as is done in the case of pure AdS$_3$ gravity in \cite{Krasnov:2000zq}. In fact, as we demonstrate in section \ref{CFT2}, the running dilaton solutions of \eqref{action} correspond to a circle reduction of pure AdS$_3$ gravity, and hence, the renormalized on-shell action \eqref{Sren} can also be obtained by a circle reduction of the Takhtajan-Zograf Liouville action obtained in \cite{Krasnov:2000zq}. In particular, $S\sbtx{global}$ corresponds to the circle reduction of the global terms in the Takhtajan-Zograf action. However, here we are interested in the local part of the on-shell action, which, as we shall see momentarily, is related to the universal local part of the Liouville action and can be expressed in terms of a Schwarzian derivative. } Like the one-point functions \eqref{1pt-fns}, \eqref{Sren} is exact in the sources $\a(t)$, $\b(t)$ and $\m(t)$, although the fact that $S\sbtx{global}$ has not been determined allows us to add an arbitrary total derivative term in this expression. In particular, successively differentiating \eqref{Sren} or the one-point functions \eqref{1pt-fns} with respect to the sources $\a(t)$, $\b(t)$ and $\m(t)$ one can evaluate any $n$-point correlation function of the operators $\ct$, $\co_\j$ and $\cj^t$ in the dual theory.

\paragraph{Effective action and the Schwarzian derivative}

As we anticipated in section \ref{model}, the sources $\a(t)$, $\b(t)$ and $\m(t)$ are locally pure gauge. In particular, the number of independent source components coincides with the number of independent local parameters of bulk diffeomorphisms and U(1) gauge transformations that preserve the Fefferman-Graham gauge \eqref{FG}, which are discussed in detail in section \ref{algebra}. Exponentiating the infinitesimal transformations \eqref{PBH-sources-I} for running dilaton solutions with respect to the local parameters $\s(t)$ and $\vf(t)$ we can express the local sources in terms of the parameters of local symmetry transformations as 
\be
\a=e^{\s}\(1+\ve'+\ve\s'\)+\co(\ve^2),\quad \b=e^{\s}\(1+\ve\s'\)+\co(\ve^2),\quad \m=\vf'+\ve'\vf'+\ve\vf''+\co(\ve^2),
\ee
where as above the primes ${}'$ denote a derivative with respect to $t$. Inserting this form of the sources in \eqref{Sren} and absorbing total derivative terms in $S\sbtx{global}$ we obtain 
\be\label{eff-action}\boxed{
S\sbtx{ren}=\frac{L}{\k_2^2}\int \tx dt\Big(\{\t,t\}-m/2\Big)+S\sbtx{global},\qquad \s=\log\t',}
\ee 
where $\t(t)$ is a `dynamical time' \cite{Engelsoy:2016xyb} and 
\be\label{Schwarzian}
\{\t,t\}=\frac{\t'''}{\t'}-\frac32\frac{\t''^2}{\t'^2},
\ee
denotes the Schwarzian derivative. This form of the effective action arises in the infrared limit of the Sachdev-Ye-Kitaev model \cite{Sachdev1993,Kitaev} and is a 
key piece of evidence for the holographic identification of this model with AdS$_2$ dilaton gravity \cite{Almheiri:2014cka,Maldacena:2016upp,Engelsoy:2016xyb}. In terms of $\s$ the action \eqref{eff-action} is just the Liouville action in one dimension, and once $S\sbtx{global}$ is taken into account, it corresponds to a circle reduction of the Takhtajan-Zograf Liouville action obtained in \cite{Krasnov:2000zq} for pure AdS$_3$ gravity. Note that it only depends on the local parameter $\s(t)$, or $\t(t)$, but not on $\ve(t)$ and $\vf(t)$. This reflects the fact that the Ward identities \eqref{WIs-I} are non-anomalous, while the trace Ward identity \eqref{trace-WI}, which corresponds to the transformation of $S\sbtx{ren}$ under  $\s(t)$ transformations, is anomalous. We therefore see that the appearance of the Schwarzian derivative is a manifestation of the conformal anomaly, exactly as in the case of conformal field theories in two dimensions.

\paragraph{Symplectic space of running dilaton solutions}

Finally, a concept that will be useful in the discussion of the conserved charges and the asymptotic symmetries in section \ref{algebra} is the symplectic form on the space of solutions. From the relations \eqref{1pt-fn-var-I} follows that for the running dilaton solutions the symplectic form is  
\be\label{symplectic-form-I}
	\O=\int \tx dt\(\d\cp_\a\wedge\d\a+\d\cp_\b\wedge\d\log\b+\cp_\m\wedge\d\m\),
\ee
where
\be\label{CV-I}
\cp_\a=\ct,\qquad \cp_\b=\a\co_\j,\qquad \cp_\m=-\a\cj^t.
\ee
This allows us to define the Poisson bracket 
\be\label{PB-I}
\{\cc_1,\cc_2\}=\int dt\(\frac{\d\cc_1}{\d\cp_\a}\frac{\d\cc_2}{\d\a}+\b\frac{\d\cc_1}{\d\cp_\b}\frac{\d\cc_2}{\d\b}+\frac{\d\cc_1}{\d\cp_\m}\frac{\d\cc_2}{\d\m}-\cc_1\leftrightarrow\cc_2\),
\ee
for any functions $\cc_{1,2}$ on the symplectic space of such solutions. 

\subsection{Holographic dictionary for constant dilaton solutions}

The holographic dictionary for constant dilaton solutions can be constructed in a similar way, except that there are a number of subtleties that require careful analysis and have often led to some confusion. The main source of these subtleties can be traced in the form of the gauge field in the solution \eqref{gen-sol-2}. As we saw in section \ref{model}, contrary to gauge fields in AdS$_{d+1}$ with $d\geq 3$, when the dilaton is constant the asymptotically leading mode for the gauge field $A_t$ is not $\wt \m(t)$, but the mode proportional to the conserved charge $Q$. The fact that this mode satisfies the conservation law in \eqref{eoms-gauge} -- in fact it is constant in two dimensions -- implies that it cannot possibly be identified with the source of a local operator in the dual theory. A similar situation arises for rank-$p$ antisymmetric tensor fields in AdS$_{d+1}$ with $p\geq d/2$, and in particular, for vector fields in AdS$_3$ \cite{thermo2}. 

In such cases there are usually two possibilities for defining the source of the dual operator. The first is to identify the source with the mode $\wt\m(t)$, which is unconstrained and possesses a gauge symmetry of the form $\wt\m(t)\to\wt\m(t) +\pa_t\wt\L(t)$. This gauge symmetry is reflected in the fact that the conjugate mode proportional to $Q$ is conserved and it is therefore naturally identified with a conserved current in the dual theory. The alternative possibility is to trivialize the constraint satisfied by the leading mode $Q$ in order to obtain an unconstrained source \cite{thermo2}. This procedure is equivalent to starting with the Hodge dual gauge field in the bulk, which has $p < d/2$. However, gauge fields in AdS$_2$ cannot be dualized and so this second option is not available in this case. It follows that  the only consistent identification of the source and one-point function dual to the gauge field $A_t$ in AdS$_2$ is that the mode $\wt \m(t)$ be identified with the source of a local current operator and the conserved mode proportional to $Q$ be identified with the conserved current. This identification is the same as for the running dilaton solutions, except that for constant dilaton solutions $\wt \m(t)$ is the asymptotically subleading mode. Nevertheless, this does not mean that $\wt \m(t)$ must necessarily be kept fixed at the boundary. Indeed, even though there are no propagating degrees of freedom, both modes are normalizable (see e.g. \cite{O'Bannon:2015gwa}) and so different boundary conditions can be imposed on AdS$_2$ gauge fields. As we shall see, this leads to different asymptotic symmetries, but also to a different holographic dual. In particular, the holographic dual with $\wt \m(t)$ identified as a source contains a local conserved current operator, while the one with $Q$ kept fixed does not.\footnote{In fact, $Q$ may be interpreted as a source for the {\em non local} Polyakov line operator $\int\tx dt\;\wt\m(t)$. Hence, depending on the boundary condition imposed on the AdS$_2$ gauge field, the dual theory possesses either a local conserved current operator or a global Polyakov line operator. In the latter case $\wt\m$ can be thought of as a dynamical gauge field in the dual theory, whose only gauge invariant observable is the Polyakov line. } 

A related subtlety arises in the derivation of the boundary counterterms that render the variational problem well posed for constant dilaton solutions. A source of confusion here is the usual folklore that the boundary counterterms are needed to remove the long distance divergences of the on-shell action. Although this is one of the properties of the boundary counterterms, it is not the fundamental property that unambiguously determines these terms. Indeed, any covariant boundary term that removes the divergences of the on-shell action is not necessarily consistent and inconsistencies can arise at the level of correlation functions. Instead, the fundamental property that determines the boundary counterterms is a well posed variational problem \cite{Papadimitriou:2005ii}. This implies that the boundary counterterms must be compatible with the symplectic structure on the space of solutions, as well as the gauge symmetries of the symplectic variables.

Preserving the symplectic structure requires that the boundary counterterms correspond to the generating function of a canonical transformation that diagonalizes the symplectic map between the phase space, parameterized by the induced fields and their conjugate momenta, and the symplectic space of asymptotic solutions, parameterized by the modes in the Fefferman-Graham asymptotic expansions \cite{Papadimitriou:2010as}. In addition, to respect the gauge symmetries of the symplectic variables, the first class constraints of the radial Hamiltonian formalism, i.e. the Ward identities of the dual theory, must be satisfied by the transformed (renormalized) canonical variables at the radial cutoff.\footnote{Failure of the renormalized variables to obey the Ward identities at the radial cutoff, in general, is an indication of quantum anomalies. However, for genuine quantum anomalies this must be a necessity rather than a poor choice of boundary counterterms.} This is clearly a stronger constraint than simply removing the divergences of the on-shell action and determines the divergent part of the boundary counterterms uniquely. Of course, a freedom of adding extra {\em finite} local counterterms, still preserving the symplectic structure and the local symmetries, remains and corresponds to the freedom of choosing a renormalization scheme. Moreover, any function of the {\em renormalized} symplectic variables can be added to the renormalized action in order to change the boundary conditions, provided the corresponding boundary conditions are admissible.    

For constant dilaton solutions of the 2D action \eqref{action} there are two types of canonical transformation that lead to the same renormalized variational principle. The first corresponds to the addition of standard local counterterms $S\sbtx{ct}[\g,A,\j]$ that depend on the gauge potential $A_t$. The resulting variation of the renormalized action takes the form 
\be
\d\(S\sbtx{reg}+S\sbtx{ct}[\g,A,\j]\)= \int \tx dt\; \(\p^t
+\frac{\d S\sbtx{ct}}{\d A_t}\)\d A_t+\cdots,
\ee
where $S\sbtx{reg}$ is the on-shell action on the radial cutoff, including the Gibbons-Hawking term, and the ellipses stand for terms involving other fields that are not important for the present argument. This implies that $S\sbtx{ct}[\g,A,\j]$ is the generating function of the canonical transformation
\be\label{canonical-1}
\(
\begin{matrix}
A_t \\ \p^t  
\end{matrix}\) \to
\(
\begin{matrix}
A_t \\ \P^t
\end{matrix}\)=\(
\begin{matrix}
A_t\\ \p^t +\frac{\d S\sbtx{ct}}{\d A_t} 
\end{matrix}\).
\ee
This type of counterterms were first considered in \cite{Castro:2008ms} and for the model \eqref{action} take the form 
\be\label{counterterms-1}
S\sbtx{ct}[\g,A,\j]=-\frac{1}{2\k_2^2L}\int \tx dt\sqrt{-\g}\(e^{-\j}-e^{-3\j}A_tA^t\).
\ee
However, there are two problems with these counterterms. The first is that the corresponding canonical transformation \eqref{canonical-1} does not diagonalize the symplectic map between the phase space variables and the modes of the constant dilaton solutions. In particular, from \eqref{gen-sol-2} we see that the leading form of $A_t$ depends on the metric mode $\wt\a$, as well as $Q$. Similarly, although $\P^t$ is asymptotically proportional to $\wt\m$, it depends on $\wt\a$ and $Q$ as well. 

The second problem is that, although, as it was shown in \cite{Castro:2008ms}, the counterterms \eqref{counterterms-1} are {\em asymptotically} invariant under local gauge transformations, the corresponding renormalized canonical variables do not respect the Ward identities on the radial cutoff. In particular, neither $\P^t$ nor $A_t$ satisfy a conservation identity reflecting the U(1) symmetry on the radial cutoff. A related observation is that, in contrast to the bare symplectic variables $\p^t$ and $A_t$, both renormalized variables $\P^t$ and $A_t$ contain the mode $\wt\m$, and so they transform non trivially under U(1) gauge transformations. The fact that the counterterms \eqref{counterterms-1} neither diagonalize the symplectic map from phase space to the space of solutions nor preserve the U(1) covariance of the symplectic variables leads us to the conclusion that these are not the correct boundary counterterms.    

A second type of canonical transformation that results in the same renormalized variational principle is generated by a boundary term of the form  
\be\label{exotic-bt}
-\int\tx dt\; \p^tA_t +\wt S\sbtx{ct}[\g,\p,\j],
\ee
where the counterterms $\wt S\sbtx{ct}[\g,\p,\j]$ are now a local functional of the canonical momentum $\p^t$ instead of the induced gauge field $A_t$. The first term in this expression performs a Legendre transform interchanging the canonical variables $A_t$ and $\p^t$, while $\wt S\sbtx{ct}$ renormalizes the gauge potential $A_t$. In particular, a variation of the total on-shell action in this case gives
\be\label{var}
\d\Big(S\sbtx{reg}-\int\tx dt\;\p^tA_t +S\sbtx{ct}[\g,\p,\j]\Big)= -\int\tx dt\(A_t
-\frac{\d S\sbtx{ct}}{\d \p^t}\)\d \p^t+\cdots,
\ee
and hence the boundary term \eqref{exotic-bt} is the generating function of the canonical transformation
\be\label{canonical-2}
\(
\begin{matrix}
A_t \\ \p^t 
\end{matrix}\) \to
\(
\begin{matrix}
-\p^t \\ A_t^{\rm ren}
\end{matrix}\)=\(
\begin{matrix}
-\p^t \\ A_t-\frac{\d S\sbtx{ct}}{\d \p^t} 
\end{matrix}\).
\ee

Note that the boundary terms \eqref{counterterms-1} and \eqref{exotic-bt} lead to the same variational principle in terms of the modes, corresponding to identifying $Q$ with the source of the dual (non local in this case) operator, and $\int \tx dt\;\wt\m$ with its one-point function. The difference is in the identification of the renormalized variables on a finite cutoff. In the former case, $A_t$ is identified with the source and $\P^t$ with the renormalized one-point function. In the latter case, $\p^t$ is identified with the source and $A_t^{\rm ren}$ with the renormalized one-point function. In both cases the counterterms are a local functional of the canonical variables identified with the source. However, the canonical transformation \eqref{canonical-2} preserves explicitly the U(1) covariance: the renormalized variables transform as their unrenormalized counterpart. In particular, the variable $\p^t$, which satisfies the U(1) Ward identity $D_t\p^t=0$, is left intact. Moreover, as we now show by computing explicitly the counterterms $\wt S\sbtx{ct}[\g,\p,\j]$, the symplectic map from phase space to the space of solutions is diagonalized.   

This example is simple enough that one can obtain the counterterms $\wt S\sbtx{ct}[\g,\p,\j]$ by using the asymptotic form of the solution \eqref{gen-sol-2} and combining the expressions \eqref{momenta} and \eqref{HJ-momenta} for the canonical momenta. The result is\footnote{This quantity vanishes identically on shell and so all the divergences of the on-shell action are canceled by the term implementing the Legendre transform in \eqref{exotic-bt}. However, $\wt S\sbtx{ct}$ is crucial for renormalizing the canonical variables. The fact that $\wt S\sbtx{ct}$, which in this case is an algebraic function of the canonical variables, vanishes on-shell indicates that constant dilaton solutions satisfy a second class constraint. As it was shown in a different context in \cite{Chemissany:2014xsa}, the boundary counterterms are ambiguous up to second class constraints, but this ambiguity is lifted to linear order in the second class constraints by requiring that the canonical variables themselves are appropriately renormalized. More generically, the renormalization of $n$-point functions determines the counterterms up to order $n$ in the second class constraints.  This second class constraint is also the reason why reducing the counterterms for 3D gravity does not give the correct boundary terms for the 2D theory in the case of constant dilaton solutions, in contrast to the running dilaton solutions.}  
\be\label{ct-constant}
\wt S\sbtx{ct}=-\frac{1}{2\k_2^2L}\int \tx dt\(\sqrt{-\g}\; e^{-\j}+ \frac{(L\k_2^2)^2}{\sqrt{-\g}}\; e^{3\j}\p^t\p_t\).
\ee
Given this form of the boundary counterterms, it is straightforward to check that the canonical transformation \eqref{canonical-2} diagonalizes the symplectic map between phase space and the space of asymptotic expansions. Namely, 
\be
\p^t\sim -\frac{1}{\k_2^2}Q,\qquad
A^{\rm ren}_t=A_t-\frac{\d S\sbtx{ct}}{\d \p^t}\sim A_t+\frac{1}{\sqrt{LQ}}\sqrt{-\g}\sim \wt\m(t),
\ee
and so the canonically transformed variables $\p^t$ and $A^{\rm ren}_t$ are directly proportional to each of the two modes in the Fefferman-Graham expansion. 

Having renormalized the symplectic variables, and hence the on-shell action, we can now address the question of boundary conditions. As we pointed out above, we may choose to keep either $\wt\m(t)$ or $Q$ fixed on the boundary, but only in the first case the dual theory has a local current operator. The variation \eqref{var}, therefore, implies that $S\sbtx{reg}-\int\tx dt\;\p^tA_t +\wt S\sbtx{ct}[\g,\p,\j]$ corresponds to the effective action of the operator dual to the gauge field $A_t$, or to the generating function of the theory defined by keeping $Q$ fixed, which does not have a local current operator. To obtain the generating function for the theory with the local current operator we need to Legendre transform back with respect to the renormalized variables. Namely, the generating functional in this case is given by 
\bal\label{Sren-II}
S\sbtx{ren}&=\lim_{u\to\infty}\Big(S\sbtx{reg}+\wt S\sbtx{ct}-\int \tx dt\; \p^t A_t+\int \tx dt\; \p^t A^{\rm ren}_t\Big)\NO\\
&=\lim_{u\to\infty}\[S\sbtx{reg}+\Big(1-\int\tx dt\; \p^t\frac{\d}{\d\p^t}\Big)\wt S\sbtx{ct}[\g,\p,\j]\],
\eal
with $\wt S\sbtx{ct}$ as in \eqref{ct-constant}.

A few remarks are in order at this point. Firstly, it is worth pointing out that if it were possible to Hodge dualize the bulk gauge field, the boundary term \eqref{exotic-bt} is exactly what one would obtain by carrying out standard holographic renormalization for the Hodge dual and then dualizing the result back. This can be demonstrated explicitly for $p$-form fields in AdS$_{d+1}$ with $p\geq d/2$ \cite{thermo2}, or even for gauge fields in non AdS spaces, as was done for the gauge field in the electric frame in \cite{An:2016fzu}. 

Secondly, the boundary term \eqref{exotic-bt}, with $\wt S\sbtx{ct}$ given in \eqref{ct-constant}, seems to be related with the counterterm for 2D dilaton gravity proposed in \cite{Grumiller:2014oha,Grumiller:2015vaa}, since the bulk fieldstrength in two dimensions only has a $ut$ component, which is proportional to the canonical momentum $\p^t$. As is the case for the boundary terms \eqref{exotic-bt}, the counterterms proposed in \cite{Grumiller:2014oha,Grumiller:2015vaa} are manifestly gauge invariant and so preserve the gauge symmetries of the symplectic variables. However, they are generically non local\footnote{We should emphasize that by ``non local'' in this case we do not mean ``non polynomial in boundary derivatives'', but rather ``non analytic''. Neither type of non locality is acceptable in the boundary counterterms.} and do not always diagonalize the symplectic map from phase space to the space of asymptotic solutions. Both of these aspects depend on the dilaton couplings in the bulk action, as well as the asymptotic solutions considered. For running dilaton solutions of the specific action \eqref{action} we have seen above that the correct counterterms are given instead by \eqref{ct-running}. Moreover, for constant dilaton solutions of the action \eqref{action}, or of any other 2D dilaton gravity theory, the counterterms in \cite{Grumiller:2014oha,Grumiller:2015vaa} are in fact non local, and so they cannot be interpreted as local counterterms in the dual theory. The reason is that $f_{\m\n}f^{\m\n}$ in eq. (5.4) of \cite{Grumiller:2015vaa}, which is proportional to $(\p^t)^2$, goes to a constant asymptotically and so, for constant dilaton, the square root cannot be expanded. This means that the boundary term in \cite{Grumiller:2015vaa} is intrinsically non local in that case. In addition, a simple calculation shows that for constant dilaton such a boundary term does not diagonalize the symplectic map between phase space and the space of asymptotic solutions, since the renormalized gauge potential $A_t^{\rm ren}$ depends not only on $\wt\m(t)$, but also on $Q$.   

However, the counterterm in \cite{Grumiller:2014oha,Grumiller:2015vaa} can give the correct result for 2D dilaton gravity, provided the dilaton potentials in the bulk action are such that there exist solutions where the modes in the asymptotic expansion of the gauge field are reversed, while at the same time the dilaton is running. The action \eqref{action}, for example, does not admit such solutions, but the action in eq. (5.3) of \cite{Grumiller:2015vaa} does. In that case, the square root in (5.4) can be expanded asymptotically for large $X$, with only the first two terms contributing to the divergences of the on-shell action, resulting in the local boundary terms \eqref{exotic-bt} and \eqref{ct-constant} here.  

In summary, for a generic 2D EMD theory, the counterterms take one of two possible forms, depending on which mode dominates asymptotically in the Fefferman-Graham expansion of the gauge field. If the conserved charge is asymptotically subleading (this can only happen in the presence of a running dilaton in two dimensions), then the boundary counterterms are of the form \eqref{ct-running}, with the function of the dilaton appearing in the counterterms determined by the dilaton couplings in the bulk action. If the conserved charge is the asymptotically leading mode in the Fefferman-Graham expansion of the gauge field, however, then the correct counterterms are of the form \eqref{exotic-bt} and \eqref{ct-constant}, with the dilaton dependence again determined by the dilaton couplings in the bulk action. It should be emphasized that the boundary counterterms that renormalize the canonical variables and cancel the divergences of the on-shell action are unambiguous and the renormalization scheme dependence does not affect them. In particular, the requirements that the counterterms diagonalize the symplectic map between phase space and the space of asymptotic solutions, and that they preserve the Ward identities on a finite cutoff,  are necessary and can be considered as the defining property of these terms. The renormalization scheme dependence corresponds to the possibility of adding extra {\em local} and {\em finite} boundary terms, that preserve the diagonalization of the symplectic map achieved by the counterterms \cite{Papadimitriou:2010as}.

\paragraph{Dual operators} Using the renormalized action \eqref{Sren-II} and the general constant dilaton solutions \eqref{gen-sol-2}, we can evaluate the exact renormalized one-point functions of the dual operators. The renormalized canonical momenta take the form  
\bsub
\label{ren-momenta-II}
\bal
\Hat\p^t_t&=\lim_{u\to\infty}e^{2u/L}\(\frac{1}{2\k^2_2}
\pa_u e^{-\j}+\frac{1}{\sqrt{-\g}}\g_{tt}\frac{\d S\sbtx{ct}}{\d\g_{tt}}\)=0,\\
\Hat\p^{t}&=\lim_{u\to\infty}\frac{e^{u/L}}{\sqrt{-\g}}\p^{t}=-\frac{1}{\k^2_2}\frac{Q}{\wt\a},\\
\Hat\p_\j&=\lim_{u\to\infty}e^{4u/L}\(-\frac{1}{\k^2_2}e^{-\j}K+\frac{1}{\sqrt{-\g}}\frac{\d S\sbtx{ct}}{\d\j}\)=\frac{2}{\k_2^2\wt L}\frac{\wt\b}{\wt\a},
\eal
\esub
and hence
\be\label{1pt-fns-const}\boxed{
\ct=2\Hat\p^t_t=0,\quad 
\co_\j=-\Hat\p_\j=-\frac{2}{\k_2^2\wt L}\frac{\wt\b}{\wt\a},\quad
\cj^t=-\Hat\p^t=\frac{1}{\k^2_2}\frac{Q}{\wt\a}.}
\ee

These one-point functions differ from those for running dilaton asymptotics in \eqref{1pt-fns} in two major ways. Firstly, the stress tensor in the theory dual to constant dilaton solutions is a trivial operator. Secondly, the scalar operator $\co_\j$ dual to the dilaton is now an irrelevant operator with scaling dimension $\wt\D_\j=2$, in agreement with e.g. \cite{Strominger:1998yg}. This is consistent with the RG flow \eqref{RG-flow} we discussed earlier, where the operator dual to the dilaton is marginally relevant with respect to the ultraviolet theory, but irrelevant with respect to the infrared one. Moreover, the general solution \eqref{gen-sol-2} does not contain a source for this irrelevant operator, since such a source would change the asymptotic form of the solution. Instead, the arbitrary function $\wt\b(t)$ in the metric corresponds to the one-point function of the irrelevant scalar operator, which can be non zero without modifying the asymptotic form of the solution. Notice that since the holographic stress tensor is identically zero for constant dilaton solutions, different values of $\wt\b(t)$ cost no energy, and so they parameterize a space of degenerate vacua, which we loosely refer to as the ``Coulomb branch'' of the theory. As we will see in section \ref{algebra}, the operator $\co_\j$ transforms anomalously under the local asymptotic symmetries, with the same conformal anomaly as in the running dilaton theory. This suggests that the microstates accounting for the black hole entropy survive in the constant dilaton solutions and should correspond to the degenerate vacua parameterized by the expectation value of $\co_\j$, i.e. $\wt\b$.

\paragraph{Ward identities} The time reparameterization and trace Ward identities are satisfied trivially in this theory because the source of the irrelevant operator is set to zero, while the charge conservation identity remains unchanged, namely 
\be\label{U1-II}
\cd_t\cj^t=0,
\ee
where $\cd_t$ now denotes the covariant derivative with respect to the boundary metric $-\wt\a^2$. Turning on a source $\wt \n(t)$ for the irrelevant operator $\co_\j$, even infinitesimally, would require the theory to be defined with a UV cutoff. To leading non trivial order in the irrelevant source, the time reparameterization and trace Ward identities take the form 
\be\label{WIs-II}
\pa_t\ct+ \co_\j\pa_t\wt\n=\co(\wt\n^2),\qquad \ct-\wt\n\co_\j=-\frac{\wt L (LQ)^{1/2}}{\k_2^2\wt\a}\pa_t\(\frac{\wt\n'}{\wt\a}\)+\co(\wt\n^2),
\ee
which are indeed trivially satisfied by the one-point functions \eqref{1pt-fns-const} in the limit $\wt\n\to0$. However, at non zero source $\wt\n$ there is again a trace anomaly that is sourced entirely by the dilaton, as in the case of running dilaton solutions. In fact, from the infrared limit \eqref{RG-flow-exp} of the RG flow between the two theories we can read off the map between the sources, namely $\wt\a=\a\b/\sqrt{LQ}$ and $\wt\n\propto \b^2$. Inserting this in the trace anomaly in \eqref{WIs-II} we recover precisely that of the running dilaton theory in \eqref{trace-WI}. Notice that the Ward identities \eqref{WIs-II} imply that the stress tensor is nonzero if and only if a source for the irrelevant scalar operator is turned on.

\paragraph{Generating functional} At zero source for the irrelevant operator $\co_\j$,
the one-point functions \eqref{1pt-fns-const} can be expressed as 
\be\label{1pt-fn-var-II}
\ct=\frac{\d S\sbtx{ren}}{\d\wt\a},\quad \co_\j=-\frac{1}{\wt\a}\frac{\d S\sbtx{ren}}{\d\wt\n},\quad \cj^t=-\frac{1}{\wt\a}\frac{\d S\sbtx{ren}}{\d\wt\m},
\ee
in terms of the renormalized on-shell action
\be\label{Sren-const}\boxed{
	S\sbtx{ren}[\wt\a,\wt\n,\wt\m]=-\frac{1}{\k_2^2\wt L}\int \tx dt\(2\wt\b\wt \n+\wt\m Q+\co(\wt\n^2)\).}
\ee
Higher order terms in the source $\wt\n$ of the irrelevant operator can be computed perturbatively by considering fluctuations around the general solution \eqref{gen-sol-2} and introducing a radial cutoff.

\paragraph{Symplectic space of constant dilaton solutions} 

As for the running dilaton solutions, the relations \eqref{1pt-fn-var-II} allows us to determine the symplectic form on the space of constant dilaton solutions:
\be\label{Omega-II}
\O=\int\tx dt\(\d\cp_{\wt\a}\wedge\d\wt\a+\d\cp_{\wt\n}\wedge\d\wt\n+\d\cp_{\wt\m}\wedge\d\wt\m\),
\ee
where
\be
\cp_{\wt\a}=\ct,\qquad \cp_{\wt\n}=-\wt\a\co_\j,\qquad\cp_{\wt\m}=-\wt\a\cj^t.
\ee
This defines the Poisson bracket
\be\label{PB-II}
\{\cc_1,\cc_2\}=\int dt\(\frac{\d\cc_1}{\d\cp_{\wt\a}}\frac{\d\cc_2}{\d\wt\a}+\frac{\d\cc_1}{\d\cp_{\wt\n}}\frac{\d\cc_2}{\d\wt\n}+\frac{\d\cc_1}{\d\cp_{\wt\m}}\frac{\d\cc_2}{\d\wt\m}-\cc_1\leftrightarrow\cc_2\),
\ee
for any functions $\cc_{1,2}$ on the symplectic space of such solutions. 

\section{3D perspective}
\label{CFT2}
\setcounter{equation}{0}

Since the action \eqref{action} can be obtained by a circle reduction from Einstein-Hilbert gravity in three dimensions, the holographic dictionaries we derived in the previous section should be consistent with the holographic dictionary for three dimensional gravity and the dual CFT$_2$. In this section we show that this is indeed the case. Moreover, as we will discuss in section \ref{3Dsymmetries}, the 3D perspective allows us to extend the asymptotic symmetry algebras, by allowing for local transformations that involve the circle direction.

The general solution of 3D Einstein-Hilbert gravity with a negative cosmological constant can be written in the Fefferman-Graham gauge
\be\label{metric_gf}
ds^2=du^2+\g_{ij}(u,x)dx^idx^j,\qquad i=1,2,
\ee
as \cite{Skenderis:1999nb}
\be\label{induced-metric}
\g_{ij}=e^{2u/L}\left(g\sub{0}_{ij}+e^{-2u/L}g\sub{2}_{ij}+e^{-4u/L}g\sub{4}_{ij}\right),
\ee
where the boundary metric $g_{(0)ij}$ is arbitrary, $g\sub{4}_{ij} = (g\sub{2}g\sub{0}^{-1}g\sub{2})_{ij}/4$ and $g\sub{2}_{ij}$ is also arbitrary except that it satisfies the constraints
\be\label{reduced-constraints}
g\sub{0}^{ij}g\sub{2}_{ij}=-\frac{L^2}{2}R[g\sub{0}],\qquad D\sub{0}^i\left(g\sub{2}_{ij}-g\sub{0}^{kl}g\sub{2}_{kl}g\sub{0}_{ij}\right)=0.
\ee
Here, $D\sub{0}_i$ denotes a covariant derivative with respect to $g\sub{0}_{ij}$. 
Defining the tensor  
\be\label{Tij}
\t_{ij}=\frac{1}{\k^2_3L}\left(g\sub{2}_{ij}-g\sub{0}^{kl}g\sub{2}_{kl}g\sub{0}_{ij}\right),
\ee
the constraints \eqref{reduced-constraints} can be written more compactly as
\be\label{2D-constraints}
D\sub{0}_i\t^i_j=0,\qquad \t^i_i=\frac{c}{24\p}R[g\sub{0}],
\ee
where 
\be\label{BH-c}
c=\frac{12\p L}{\k^2_3}=\frac{3L}{2G_3},
\ee
is the Brown-Henneaux central charge \cite{Brown:1986nw}. This form of the constraints allows us to identify $\t_{ij}$ with the stress tensor in the dual CFT$_2$. In terms of $\t_{ij}$ the metric \eqref{metric_gf}-\eqref{induced-metric} becomes
\begin{align}
	\label{3D-metric}
	\begin{aligned}
		ds^2&=du^2+e^{2u/L}\left\{g\sub{0}_{ij}+2e^{-2u/L}\left(\frac{\k^2_3L}{2}\t_{ij}-\frac{L^2}{4}
		R[g\sub{0}]g\sub{0}_{ij}\right)\right.\\
		&\hspace{0.5cm}\left.+e^{-4u/L}\left(\frac{\k^2_3L}{2}\t_{ik}-\frac{L^2}{4} R[g\sub{0}]g\sub{0}_{ik}\right)
		\left(\frac{\k^2_3L}{2}\t^k_j-\frac{L^2}{4} R[g\sub{0}]\d^k_j\right)\right\}dx^idx^j.
	\end{aligned}
\end{align}

\paragraph{Kaluza-Klein reduction to 2D} To make contact with the solutions of the EMD theory in 2D we parameterize the AdS$_3$ coordinates as $x^i=\{u,t,z\}$, where $x^a=\{u,t\}$ cover the AdS$_2$ subspace and $z$ is periodically identified as $z\sim z+R_z$ with period $R_z$. Using the Kaluza-Klein Ansatz
\cite{Strominger:1998yg,Castro:2014ima}
\be
ds_3^2=e^{-2\j}\(dz+A_a dx^a\)^2+g_{ab}dx^a dx^b=du^2+\g_{tt}dt^2+e^{-2\j}\(dz+A_t dt\)^2,
\ee
leads to the following relations between the metric in 3D and the various fields of the EMD theory in 2D:
\be\label{KK-2D}
\g^{(3)}_{tt}=\g_{tt}+e^{-2\j}A_t^2,\qquad
\g^{(3)}_{tz}=e^{-2\j}A_t,\qquad
\g^{(3)}_{zz}=e^{-2\j}.
\ee
Moreover, the gravitational constants are related as
\be
\k_3^2=R_z\k_2^2.
\ee

\paragraph{Solving the constraints via the Liouville equation}

To fully specify the metric \eqref{3D-metric} it is necessary to solve the constraints \eqref{2D-constraints} so that the stress tensor $\t_{ij}$ is expressed as a functional of the boundary metric $g\sub{0}_{ij}$. This can be achieved with the help of an auxiliary Liouville field.

It is straightforward to check that a stress tensor of the form
\be\label{Liouville-redef}
\t_{ij}=\frac{2}{q^2}e^{\frac{q}{2}\vf_p}\left(D\sub{0}_i D\sub{0}_j
-\frac12g\sub{0}_{ij}\square\sub{0}\right)e^{-\frac q 2\vf_p}+\frac{1}{2q^2}R[g\sub{0}]g\sub{0}_{ij},
\ee
where $1/q^2=c/24\p$ is proportional to the Brown-Henneaux central charge \eqref{BH-c} and the auxiliary scalar field $\vf_p$ satisfies the Liouville equation
\be\label{liouville-eq}
q\square_{(0)}\vf_p-p e^{q\vf_p}=R[g_{(0)}],
\ee
with some arbitrary parameter $p$, automatically satisfies the constraints \eqref{2D-constraints}. Conversely, provided $\t_{ij}$ satisfies the constraints \eqref{2D-constraints}, any solution of the linear equation 
\be\label{liouville-linear}
\left(\t_{ij}-\frac{1}{2q^2}R[g\sub{0}]g_{(0)ij}-\frac{2}{q^2}\Big(D_{(0)i}D_{(0)j}
-\frac12g_{(0)ij}\square_{(0)}\Big)\right)e^{-\frac{q}{2}\vf}=0,
\ee
satisfies the Liouville equation \eqref{liouville-eq} for some parameter $p$. This last statement can be proven by multiplying \eqref{liouville-linear} with
$e^{\frac{q}{2}\vf}$, then taking the covariant divergence with respect to one of the indices, and then multiplying by $e^{-q\vf}$. This gives
\be
\pa_i\left(e^{-q\vf}(R[g]-q\square_{(0)}\vf)\right)=0,
\ee
which is the Liouville equation \eqref{liouville-eq} with the parameter
$p$ emerging as an integration constant. 

This result implies that the general solution of the constraints \eqref{2D-constraints} can be parameterized as a Liouville stress tensor of the form \eqref{Liouville-redef}, in terms of the general solution of the family of Liouville equations \eqref{liouville-eq}. As we shall show below, this observation plays a crucial role in the relation between the holographic dictionaries for the EMD theory \eqref{action} in 2D and that for 3D Einstein-Hilbert gravity. 

\subsection{Running dilaton solutions from spacelike reduction}

Using the Kaluza-Klein relations \eqref{KK-2D}, the general solution \eqref{gen-sol} can be uplifted to three dimensions. In particular, we read off the following boundary metric $g\sub{0}_{ij}$ and CFT$_2$ stress tensor $\t_{ij}$ that parameterize the metric \eqref{3D-metric} in three dimensions: 
\be\label{CFT-metric-I}
g_{(0)zz}=\b^2,\qquad
g_{(0)zt}=\b^2\m,\qquad
g_{(0)tt}=-(\a^2-\b^2\m^2),
\ee
and
\bsub\label{CFT-tensor-I}
\bal
R_z\t_{zz}&=\b\co_\j,\\
R_z\t_{zt}&=\b\m\co_\j+\frac{\a^2}{\b}\cj^t,\\
R_z\t_{tt}&=-\frac{\a^2}{\b}\ct+\b\m^2\co_\j+\frac{2\a^2\m}{\b}\cj^t.
\eal
\esub
Notice that the component $g_{(0)zz}$ of the boundary metric in three dimensions is positive definite and so the running dilaton solutions of the 2D EMD theory are obtained through a spacelike circle reduction of 3D Einstein-Hilbert gravity.

\paragraph{Ward identities} The Ricci scalar of the boundary metric \eqref{CFT-metric-I} is
\be
R[g_{(0)}]=2\(\frac{\b''}{\a^2\b}-\frac{\a'\b'}{\a^3\b}\).
\ee
Evaluating the trace of the stress tensor $\t_{ij}$ through the relations \eqref{CFT-tensor-I} then reproduces the trace constraint in  \eqref{2D-constraints}, namely
\be
R_z \t^i_i=\frac{1}{\b}\(\ct+\co_\j\)= \frac{R_zL}{\k_3^2}\(\frac{\b''}{\a^2\b}-\frac{\a'\b'}{\a^3\b}\)=\frac1\b\ca,
\ee
where $\ca$ is the conformal anomaly of the 2D EMD theory with running dilaton defined in \eqref{trace-WI}. Hence, the conformal anomaly of the 2D theory, which is entirely due to the running dilaton, precisely matches the metric conformal anomaly in the 2D CFT. Moreover, it is straightforward to check that the relations \eqref{CFT-tensor-I} map the Ward identities \eqref{WIs-I} to the conservation of the stress tensor $\t_{ij}$ in \eqref{2D-constraints}.

\paragraph{Relation to the Liouville equation} We now show that the expressions \eqref{CFT-tensor-I} for the stress tensor can be written in the form \eqref{Liouville-redef}, in terms of a specific solution of the Liouville equation in the background metric \eqref{CFT-metric-I}, which takes the form
\be
\frac{q}{\a\b}\[\frac{\a^2-\b^2\m^2}{\a\b}\pa_z^2\vf_p+\frac{\b\m}{\a}\pa_t\pa_z\vf_p+\pa_t\(\frac{\b\m}{\a}\pa_z\vf_p-\frac{\b}{\a}\pa_t\vf_p\)\]-p e^{q\vf_p}=\frac{2}{\a\b}\pa_t\(\frac{\b'}{\a}\).
\ee
For $p=0$ a solution of this equation is 
\be\label{Liouville-sol-I}
\vf_0(t,z)=c_0 z+h(t),\quad h'(t)=c_0\m+c_1\frac{\a}{\b}-\frac{2\b'}{q\b},
\ee
where $c_0$ and $c_1$ are arbitrary constants. Inserting this solution in the expression \eqref{Liouville-redef} for the stress tensor we obtain
\bsub
\label{Liouville-I}
\bal
\t_{zz}&=\frac{1}{q^2}\(\frac{q^2}{4}\(c_0^2+c_1^2\)-\frac{\b'^2}{\a^2}+\frac{2\b\b''}{\a^2}-\frac{2\b\b'\a'}{\a^3}\),\\
\t_{zt}&=\frac{c_0c_1\a}{2\b}+\m\t_{zz},\\
\t_{tt}&=\frac{1}{q^2}\(\frac{q^2}{4}(c_0^2+c_1^2)\frac{\a^2}{\b^2}-\frac{\b'^2}{\b^2}\)+\frac{c_0c_1\a\m}{\b}+\m^2\t_{zz},
\eal
\esub
which coincide with the expressions \eqref{CFT-tensor-I} with the identifications
\be
m=\frac{\k_2^2R_z}{2L}(c_0^2+c_1^2),\qquad Q=\frac{\k_2^2 R_z}{2}c_0c_1,\qquad \frac{1}{q^2}=\frac{c}{24\p}=\frac{L}{2\k_2^2 R_z}.
\ee

\subsection{Constant dilaton solutions from null reduction}

Uplifting the constant dilaton solutions \eqref{gen-sol-2} using the Kaluza-Klein relations \eqref{KK-2D} results in the following boundary metric $g\sub{0}_{ij}$ and CFT$_2$ stress tensor $\t_{ij}$: 
\be\label{CFT-metric-II}
g_{(0)zz}=0,\qquad
g_{(0)zt}=-\sqrt{LQ}\;\wt\a,\qquad
g_{(0)tt}=-2\sqrt{LQ}\; \wt\a\wt\m,
\ee
and
\bsub\label{CFT-tensor-II}
\bal
\k_3^2\t_{zz}&=Q=\k_2^2\wt\a\cj^t,\\
\k_3^2\t_{zt}&=Q\wt\m =\k_2^2\wt\a\wt\m\cj^t,\\
\k_3^2\t_{tt}&=-\frac{2\wt\a\wt\b}{\sqrt{LQ}\;\wt L}+Q\wt\m^2=\k_2^2\wt\a\(\frac{\wt\a}{\sqrt{QL}}\co_\j+\wt\m^2\cj^t\).
\eal
\esub
Since the component $g_{(0)zz}$ of the boundary metric vanishes, the constant dilaton solutions of 2D EMD theory are obtained by a null reduction of 3D Einstein-Hilbert gravity \cite{Strominger:1998yg,Castro:2014ima}. Note that although AdS$_2$ in this case has radius $\wt L=L/2$, the corresponding AdS$_3$ radius is still $L$.

It was pointed out in \cite{Castro:2014ima} that this form of the 2D CFT stress tensor implies that the constant dilaton solutions are compatible with 
the Comp\`ere-Song-Strominger (CSS) boundary conditions for 3D gravity found in \cite{Compere:2013bya}. Indeed, setting 
\be
\wt\a =\frac{1}{2\sqrt{LQ}},\quad \wt\m=-P'(t),\quad Q=\frac{\k_3^2}{2\p }\D,\quad \wt\b=-\frac{L^2Q\k_3^2}{4\p}L\sbtx{CSS}(t),
\ee
where $P$, $\D$ and $L\sbtx{CSS}$ refer to notation used in \cite{Compere:2013bya}, reproduces exactly the boundary metric and stress tensor in \cite{Compere:2013bya}. 
However, more than one boundary conditions within this space of asymptotic solutions are compatible with the symplectic structure \eqref{Omega-II}. Since $\wt\n$ is an infinitesimal source that must be set to zero in the solutions, any choice of boundary conditions must include keeping $\wt\n$ fixed. One choice of consistent boundary conditions is the CSS boundary conditions \cite{Compere:2013bya}, which correspond to
\be\label{CSS-bc}
\d\wt\n=0,\quad \d Q=0,\quad \d\wt\a=0,
\ee
with $\wt\m$ and $\wt\b$ arbitrary dynamical variables. A second class of consistent boundary conditions on these solutions is
\be\label{new-bc}
\d\wt\n=0,\quad \d\wt\m=0,\quad \d\wt\a=0,
\ee  
with $\wt\b$ and $Q$ arbitrary. These two choices of boundary conditions define different dual theories, with different operators and different symmetries. In particular, the theory corresponding to CSS boundary conditions \eqref{CSS-bc} does not have a local current operator, while the one dual to the boundary conditions \eqref{new-bc} does. We will consider the asymptotic symmetries and corresponding conserved charges for both boundary conditions in sections \ref{algebra} and \ref{3Dsymmetries}.  

Notice that, in terms of the boundary metric \eqref{CFT-metric-II}, the boundary conditions \eqref{new-bc} do not correspond to keeping the Weyl factor fixed, since the charge $Q$ is allowed to vary. This is completely consistent with the symplectic structure and the variational problem, but shows that the standard Dirichlet boundary conditions on the space of constant dilaton solutions of the 2D EMD theory uplift to generalized Dirichlet boundary conditions in 3D. Such generalized Dirichlet boundary conditions arise naturally in asymptotically AdS spaces since the bulk fields do not induce fields on the conformal boundary, but rather a conformal class of fields, i.e. a set of sources defined up to local Weyl rescalings, and are well defined provided the conformal anomaly vanishes \cite{Papadimitriou:2005ii}. This suggests that both boundary conditions \eqref{CSS-bc} and \eqref{new-bc} can be extended by allowing $\wt\a(t)$ to vary, in parallel with the generalization of the Brown-Henneaux boundary conditions in \cite{Troessaert:2013fma}. However, we will not discuss this type of boundary conditions further here.  

\paragraph{Ward identities} As we saw above, with zero source for the irrelevant scalar operator, the Ward identities for constant dilaton boundary conditions are satisfied trivially. Using the relations \eqref{CFT-metric-II} and \eqref{CFT-tensor-II} these Ward identities can be expressed in the form \eqref{2D-constraints}, or more explicitly,  
\be
\pa_t\(\wt\a\t^t_z\)=0,\qquad \pa_t\(\t^t_t-\wt\m\t^t_z\)=0,\qquad \t^t_t+\t^z_z=0,
\ee
reflecting respectively charge conservation, time reparameterizations, and radial reparameterizations. Although these are trivially satisfied by the solutions \eqref{CFT-metric-II} and \eqref{CFT-tensor-II}, the form of the conserved quantities is useful in order to define the conserved charges in section \ref{algebra}.

\paragraph{Relation to the Liouville equation} Finally, let us determine the solution of the Liouville equation \eqref{liouville-eq} that gives rise to the 
stress tensor \eqref{CFT-tensor-II} through the identity \eqref{Liouville-redef}. In the background metric \eqref{CFT-metric-II} the Liouville equation \eqref{liouville-eq} takes the form
\be
\frac{2q}{\sqrt{LQ}\;\wt\a}\(\wt\m\pa_z-\pa_t\)\pa_z\vf_p-p e^{q\vf_p}=0.
\ee
For $p=0$ the general solution of this equation is
\be
q\;\vf_0(t,z)=\log\pa_+\cf(x^+)+\log\pa_-\cg(x^-),
\ee
where $\cf$ and $\cg$ are arbitrary functions of their arguments and $\pa_\pm$ denote derivatives with respect to the variables
\be\label{liouville-coords}
x^+=2\sqrt{LQ}\int^t\tx dt' \wt\a(t'),\qquad x^-=z+\int^t \tx dt'\wt\m(t').
\ee
Note that in terms of $x^\pm$ the boundary metric \eqref{CFT-metric-II} takes the canonical form 
\be
ds_{(0)}^2=-dx^+dx^-.
\ee 
The stress tensor \eqref{Liouville-redef} then becomes 
\bsub\label{Liouville-tensor-II}
\bal
\t_{zz}&=-\frac{1}{q^2}\(\pa_-\(\frac{\pa_-^2\cg}{\pa_-\cg}\)-\frac{(\pa_-^2\cg)^2}{2(\pa_-\cg)^2}\),\\
\t_{zt}&=\wt\m \t_{zz},\\
\t_{tt}&=-\frac{4LQ\wt\a^2}{q^2}\(\pa_+\(\frac{\pa_+^2\cf}{\pa_+\cf}\)-\frac{(\pa_+^2\cf)^2}{2(\pa_+\cf)^2}\)+\wt\m^2\t_{zz},
\eal
\esub
which involves the Schwarzian derivatives \eqref{Schwarzian} of $\cg$ with respect to $x^-$ and of $\cf$ with respect to $x^+$. These expressions for the stress tensor of the 2D CFT matches the expressions \eqref{CFT-tensor-II} obtained from the 2D EMD theory provided we identify     
\be\label{Liouville-sol-II}
\pa_-\cg={\rm sech}^2\(\sqrt{\frac{Q}{L}}\;x^-\),\qquad
\wt\b=\frac{L^2}{2}(LQ)^{3/2}\;\wt\a\(\pa_+\(\frac{\pa_+^2\cf}{\pa_+\cf}\)-\frac{(\pa_+^2\cf)^2}{2(\pa_+\cf)^2}\),
\ee
with $\cf(x^+)$ arbitrary.

\section{Conserved charges and asymptotic symmetry algebras}
\label{algebra}
\setcounter{equation}{0}

In this section we identify the asymptotic symmetries preserved by the running and constant dilaton solutions of the 2D theory \eqref{action}, and construct the associated conserved charges. It should be emphasized that the symmetries preserved depend fundamentally on the identification of the sources in the dual theory. Hence, even within the same space of asymptotic solutions, identifying different modes as sources leads to different asymptotic symmetries. Here we consider standard Dirichlet boundary conditions for the running dilaton solutions, but for the constant dilaton solutions we consider both Dirichlet and CSS boundary conditions. 

In order to identify the asymptotic symmetries we first need to know how the modes parameterizing the solutions transform under the local bulk transformations that preserve the Fefferman-Graham gauge. Bulk diffeomorphisms with this property are known as Penrose-Brown-Henneaux (PBH) diffeomorphisms \cite{Penrose,Brown:1986nw} (see also \cite{Imbimbo:1999bj}), but they can be generalized to other local symmetries, such as gauge transformations. In general, the local bulk transformations preserving the Fefferman-Graham gauge mix, and so we will refer to all such transformations collectively as generalized PBH transformations.

Under an infinitesimal bulk diffeomorphism the fields of the 2D model \eqref{action} transform as 
\bsub
\bal
&\d_\x g_{uu} =\cl_\x g_{uu}=\dot\x^u,\quad
\d_\x g_{tt} = \cl_\x g_{ut}=\g_{tt}(\dot\x^t+\pa^t\x^u),\quad
\d_\x g_{tt} = \cl_\x g_{tt}=L_\x\g_{tt}+2K_{tt}\x^u,\\
&\d_\x A_{u} =\cl_\x A^\L_{u}=\dot\x^{t}A_{t},\quad
\d_\x A_{t} =\cl_\x A_{t}=L_\x A_{t}+\x^u\dot  A_{t},\quad
\d_\x \cl_\x\j= \cl_\x\j =L_\x\j+\x^u\dot\j,
\eal
\esub
where $\cl_\x$ denotes the Lie derivative with respect to the vector $\x^a$, while $L_\x$ stands for the Lie derivative with respect to the transverse component $\x^t$. Moreover, under an infinitesimal gauge transformation the gauge field transforms as
\be
\d_\L A_{u}=\dot \L,\qquad \d_\L A_{t}=\pa_t\L.
\ee
To preserve the Fefferman-Graham gauge \eqref{gauge-fixing}, therefore, we must demand 
\be
\cl_\x g_{uu}=\cl_\x g_{ut}=0,\quad (\cl_\x+\d_\L)A_{u}=0.
\ee
This leads to a set of equations for the parameters $\x^a(u,t)$ and $\L(u,t)$, with  general solution  \cite{Papadimitriou:2005ii}
\be\label{gen-PBH}
\x^u=\s(t),\quad
\x^t=\ve(t)+\pa_t\s(t)\int_u^\infty du'\g^{tt}(u',t),\quad
\L=\vf(t)+\pa^{t}\s(t)\int_u^\infty du'A_{t}(u',t), 
\ee
where $\ve(t)$, $\s(t)$ and $\vf(t)$ are arbitrary functions of time. They correspond respectively to time reparameterizations, i.e. boundary diffeomorphisms, Weyl and gauge transformations. Under these transformations the dynamical fields transform as 
\be\label{PBH-transformations}
\d_\x \g_{tt}=L_\x\g_{tt}+2K_{tt}\x^u,\quad
(\cl_\x+\d_\L) A_{t}=L_\x A_{t}+\x^u\dot  A_{t}+\pa_t\L,\quad
\d_\x\j=L_\x\j+\x^u\dot\j.
\ee
Inserting the asymptotic expansions of these fields in the general solution \eqref{gen-PBH} determines the asymptotic form of PBH transformations. Using the resulting form of the PBH transformations, together with the asymptotic expansions of the fields in \eqref{PBH-transformations} determines how the modes transform.

Finally, defining the generalized Lie bracket for diffeomorphisms and gauge transformations\footnote{This is a well defined Lie bracket since it satisfies the Jacobi identity. Moreover, in the present setting in can be derived by Kaluza-Klein reduction from 3D pure gravity PBH transformations.} 
\bsub
\label{PBH-commutators}
\bal
[\x_1,\x_2]&:=\d_{\x_1}\x_2=\cl_{\x_1}\x_2,\\
[\x,\L]&:=\d_{\x}\L-\d_\L\x=\d_{\x}\L,\\
[\L,\x]&:=\d_\L\x-\d_{\x}\L=-\d_{\x}\L,\\
[\L_1,\L_2]&:=\d_{\L_1}\L_2-\d_{\L_2}\L_1,
\eal
\esub
one can compute the algebra of local transformations preserving the Fefferman-Graham gauge. Strictly speaking, though, this is not an algebra since it closes up to field dependent transformation parameters. However, as we shall see, when restricted to asymptotic symmetries this becomes a proper algebra: the asymptotic symmetry algebra.

\subsection{Running dilaton solutions}

Inserting the general running dilaton solution \eqref{gen-sol} in \eqref{gen-PBH} and 
\eqref{PBH-transformations} we find that PBH transformations in this case act on the sources as 
\be\label{PBH-sources-I}
\d\sbtx{PBH}\a =\pa_t(\ve\a)+\a\s/L,\qquad
\d\sbtx{PBH}\b =\ve\b'+\b\s/L,\qquad
\d\sbtx{PBH}\m =\pa_t(\ve\m+\vf),
\ee
and on the one-point functions as
\bsub\label{PBH-1pt-fns-I}
\bal
\d\sbtx{PBH}\ct &=\ve\pa_t\ct-\frac{\s}{L}\ct+\frac{\b'\s'}{\k_2^2\a^2},\\
\d\sbtx{PBH}\co_\j &=\ve\pa_t\co_\j-\frac{\s}{L}\co_\j+\frac{L\ve}{2\k_2^2\b}\pa_t\(\frac{\b'^2}{\a^2}\)+\frac{L}{\k_2^2\a}\pa_t\(\frac{\b\s'}{L\a}\),\\
\d\sbtx{PBH}\cj^t &=-\(\frac{\pa_t(\ve\a)}{\a}+\frac{\s}{L}\)\cj^t.
\eal
\esub

These transformations are intimately connected with the Ward identities \eqref{WIs-I} and \eqref{trace-WI}. Rewriting these identities in terms of the symplectic variables \eqref{CV-I} and defining the function\footnote{In order for the functional derivatives to be well defined one must add a `boundary' term to this function on the initial and final times. Such terms are related to the conserved charges. However, we will determine the conserved charges by an alternative argument and so we do not show explicitly these boundary terms.} 
\be
\bb H[\ve,\s,\vf]=\int \tx dt\(-\ve\(\a\pa_t \cp_{\a}-\frac{\b'}{\b} \cp_{\b}+\m\pa_t\cp_{\m}\)+\frac{\s}{L}\(\a\cp_{\a}+\cp_{\b}-\a\ca\)-\vf\pa_t\cp_{\m}\),
\ee
on the space of running dilaton solutions, one can check that 
\be
\d\sbtx{PBH}X=\{\bb H[\ve,\s,\vf],X\},
\ee
where $\{\cdot,\cdot\}$ denotes the Poisson bracket \eqref{PB-I} and $X$ stands for any of the canonical variables. It follows that PBH transformations are generated by the Ward identities. Conversely, the transformation of the renormalized action under arbitrary PBH transformation (invariant up to anomalies) leads to an alternative derivation of the Ward identities.  

\paragraph{Asymptotic symmetries} Asymptotic symmetries correspond to PBH transformations that leave the sources invariant, i.e.\footnote{As we mentioned earlier, this condition can be relaxed e.g. by demanding that the sources are preserved up to a Weyl transformation as in \cite{Troessaert:2013fma}, but we will not consider such generalized boundary conditions here.} 
\be\label{strict-CKV}
\d\sbtx{PBH}\(\text{sources}\)=0.
\ee
The solutions of this condition correspond to boundary conformal Killing vectors (BCKV) and are in one-to-one correspondence with {\em asymptotic} bulk Killing vectors \cite{Papadimitriou:2005ii}, where the qualification `asymptotic' means that they are not necessarily symmetries of the one-point functions. Generically, if the conformal anomaly is numerically non zero, then only boundary Killing vectors lead to conserved charges. However, in certain cases the form of the anomaly allows one to define conserved charges even for boundary conformal Killing vectors that are not Killing \cite{thermo2}. As we will see, the conformal anomaly that arises in the present context is one of these cases.

The general solution of the condition \eqref{strict-CKV} for the transformations
\eqref{PBH-sources-I} is
\be\label{symmetries-I}
\ve=\x_1\frac{\b}{\a},\qquad \s/L=-\x_1\frac{\b'}{\a},\qquad \vf=\x_2-\x_1\frac{\b}{\a}\m,
\ee
where $\x_{1,2}$ are arbitrary constants. The symmetry algebra is therefore  $u(1)\oplus u(1)$, whose corresponding charges are the mass and the electric charge. The one-point functions are strictly invariant under these global transformations and so this algebra is preserved on the Hilbert space of the dual theory, i.e. there is no anomaly.

\paragraph{Conserved charges} A simple and general way to derive the conserved charges is to consider the variation of the renormalized action under PBH transformations. From the relations \eqref{1pt-fn-var-I} between the renormalized on-shell action and the one-point functions follows that
\be\label{PBH-var}
\d\sbtx{PBH} S\sbtx{ren}=\int \tx dt\Big(\ct\d\sbtx{PBH}\a+\frac{\a}{\b}\co_\j\d\sbtx{PBH}\b-\a\cj^t\d\sbtx{PBH}\m\Big).
\ee
When evaluated on generic PBH transformation \eqref{PBH-sources-I} this variation is equal to 
\be
\d\sbtx{PBH} S\sbtx{ren}=\frac1L\int \tx dt\;\a\s\;\ca,
\ee
where $\ca$ is the conformal anomaly. As we pointed out above, inserting the PBH transformation \eqref{PBH-sources-I} of the sources and using the fact that $\ve(t)$, $\s(t)$ and $\m(t)$ are arbitrary functions leads to an alternative derivation of the Ward identities \eqref{WIs-I} and \eqref{trace-WI}. 

When evaluated instead on the boundary conformal Killing vectors \eqref{symmetries-I} instead, the variation \eqref{PBH-var} gives identically zero. Using the Ward identities and keeping the total derivative terms leads to the general expressions for the corresponding conserved charges. In particular, for the symmetry transformations \eqref{symmetries-I} the variation of the renormalized action takes the form 
\bal
0=\d\sbtx{BCKV} S\sbtx{ren}&=\int \tx dt\;\pa_t\(\x_1\(\b\ct-\frac{L}{2\k_2^2}\frac{\b'^2}{\a^2}\)-\x_2\a\cj^t\),
\eal
which leads to the two commuting conserved charges
\be\label{2D-charges-I}
\cq_1=-\(\b\ct-\frac{L}{2\k_2^2}\frac{\b'^2}{\a^2}\)=\frac{mL}{2\k_2^2},\qquad
\cq_2=\a\cj^t=\frac{Q}{\k_2^2}.
\ee
These two conserved charges carry a representation of the asymptotic symmetry algebra $u(1)\oplus u(1)$ on the Hilbert space of the dual 1D theory.

\subsection{Constant dilaton solutions} 

Inserting the general constant dilaton solution \eqref{gen-sol-2} in \eqref{gen-PBH} and 
\eqref{PBH-transformations} we find that PBH transformations act on the sources as 
\be\label{PBH-sources-II}
\d\sbtx{PBH}\wt\a =\pa_t(\ve\wt\a)+\wt\a\s/\wt L+\co(\wt\n),\qquad
\d\sbtx{PBH}\wt\n =\ve\wt\n'+\wt\n\s/\wt L,\qquad
\d\sbtx{PBH}\wt\m =\pa_t(\ve\wt\m+\vf),
\ee
and on the one-point functions as
\bsub\label{PBH-1pt-fns-II}
\bal
\d\sbtx{PBH}\ct &=-\frac{\sqrt{LQ}}{\k_2^2}\frac{\wt\n'}{\wt\a^2}\s',\\
\d\sbtx{PBH}\co_\j &=\ve\pa_t\co_\j-\frac{2\s}{\wt L}\co_\j+\frac{\sqrt{LQ}}{\k_2^2\wt\a}\pa_t\(\frac{\s'}{\wt\a}\),\\
\d\sbtx{PBH}\cj^t &=-\(\frac{\pa_t(\ve\wt\a)}{\wt\a}+\frac{\s}{\wt L}\)\cj^t.
\eal
\esub
Note that we have included the infinitesimal source $\wt\n$ of the irrelevant scalar operator $\co_\j$ in these transformations since it is required in order to generate the PBH transformations through the Poisson bracket. In particular, defining the function 
\be
\bb H[\ve,\s,\vf]=\int \tx dt\Big(-\ve\(\wt\a\pa_t \cp_{\wt\a}-\wt\n' \cp_{\wt\n}+\wt\m\pa_t\cp_{\wt\m}\)+\frac{\s}{\wt L}\(\wt\a\cp_{\wt\a}+\wt\n\cp_{\wt\n}-\wt\a\ca\)-\vf\pa_t\cp_{\wt\m}\Big),
\ee
on phase space, corresponding to the Ward identities \eqref{U1-II} and \eqref{WIs-II}, one can check that the PBH transformations are generated by the Poisson bracket \eqref{PB-II} according to 
\be
\d\sbtx{PBH}X=\{\bb H[\ve,\s,\vf],X\},
\ee
where again $X$ stands for any of the symplectic variables.

\paragraph{Asymptotic symmetries for Dirichlet boundary conditions} The asymptotic symmetries again correspond to the subset of PBH transformations that leave the sources invariant. However, for the constant dilaton solutions we want to consider two different boundary conditions on the gauge field, which lead to different asymptotic symmetries. For Dirichlet boundary conditions, corresponding to keeping $\wt\m(t)$ fixed,\footnote{Note that we call these Dirichlet boundary conditions even though the mode $\wt\m$ is asymptotically subleading. This is a matter of terminology to some extent, but it is motivated by the fact that such boundary conditions are indeed Dirichlet once uplifted to 3D. It is also natural to treat them as Dirichlet in the context of $p$-form antisymmetric tensors \cite{thermo2}.} the subset of PBH transformations that leave the sources invariant is 
\be\label{symmetries-II}
\ve(t)=\frac{\z}{2\sqrt{LQ}\wt\a} ,\qquad \s(t)=-\frac{\wt L}{2\sqrt{LQ}\wt\a}\z',\qquad \vf=-\frac{\z}{2\sqrt{LQ}\wt\a}\wt\m+\x_2,
 \ee
where $\z(t)$ is an arbitrary function of time, $\x_2$ an arbitrary constant, and the normalization of $\z$ has been chosen for reasons that will become apparent momentarily. Note that these are the asymptotic symmetries when the source $\wt\n$ of the irrelevant operator $\co_\j$ is set to zero. This is precisely the reason why the solution \eqref{symmetries-II} contains an arbitrary function of time. Once this irrelevant source is turned on the asymptotic symmetries are broken down to those for running dilaton solutions, namely \eqref{symmetries-I}. 

The symmetry algebra of the BCKVs \eqref{symmetries-II} is ${\rm Witt}\oplus u(1)$, where the Witt algebra is the classical Virasoro algebra, i.e. with zero central charge. However, not all one-point functions remain invariant under this asymptotic symmetries, which implies that the symmetry is broken by anomalies in the Hilbert space of the dual 1D theory. In particular, inserting the BCKVs \eqref{symmetries-II} in the transformation \eqref{PBH-1pt-fns-II} of the one-point functions we find that $\ct$ and $\cj^t$ are invariant, while the irrelevant scalar operator transforms as 
\be\label{anomalous-scalar}
\d_\z\co_\j =\z\pa_+\co_\j+2(\pa_+\z)\co_\j-\frac{2L (LQ)^{3/2}}{\k_2^2}\pa_+^3\z,
\ee
where $x^+$ is the null coordinate defined in \eqref{liouville-coords}. We immediately recognize this expression as the anomalous transformation of a Virasoro current. This should come as no surprise, given that we saw earlier in \eqref{Liouville-sol-II} that the relation with the Liouville equation implies that $\co_\j$ is proportional to a Schwarzian derivative with respect to $x^+$. In fact, the coefficient of the anomalous term in \eqref{anomalous-scalar} is precisely such that the time-time component of the corresponding stress tensor on the boundary of AdS$_3$, given in \eqref{CFT-tensor-II}, transforms with the Brown-Henneaux central charge, in agreement with the Liouville expression \eqref{Liouville-tensor-II}.    

The anomalous transformation \eqref{anomalous-scalar} of $\co_\j$ is a manifestation of the trace anomaly in \eqref{WIs-II} and it implies that the Witt algebra is only an asymptotic symmetry, which is broken to the global $sl(2,\bb R)$ subalgebra in the interior of AdS$_2$. In holographic terms, the Witt algebra is broken by the conformal anomaly to $sl(2,\bb R)$ on the Hilbert space of the dual theory. This is analogous to the breaking of the Virasoro symmetry in any 2D CFT by the stress tensor. However, as we show next, in the case of conformal quantum mechanics, the Witt algebra is trivially realized on the Hilbert space and it does not extend to a Virasoro algebra.

\paragraph{Conserved charges for Dirichlet boundary conditions} Applying the symmetry transformations \eqref{symmetries-II} to the variation of the renormalized on-shell action, namely
\be
0=\d\sbtx{BCKV} S\sbtx{ren}=\int \tx dt\(\ct\d\sbtx{BCKV}\wt\a-\wt\a\co_\j\d\sbtx{BCKV}\wt\n-\wt\a\cj^t\d\sbtx{BCKV}\wt\m\),
\ee
and keeping the total derivative terms leads to the conserved charges 
\be
\cq[\ve]=\wt\a\ct\ve(t)=0,\qquad \cq=\wt\a\cj^t=\frac{Q}{\k_2^2}.
\ee
In particular, the conserved charges associated with the Witt algebra generators vanish identically, and so the conformal algebra is realized trivially on the Hilbert space of conformal quantum mechanics. Hence, although the asymptotic symmetry algebra is larger than that for the running dilaton solutions, the representation of the symmetry algebra on the Hilbert space is trivial, except for a global $u(1)$.

\paragraph{Asymptotic symmetries for CSS boundary conditions}

Let us now consider the alternative CSS boundary conditions. These correspond to keeping $Q$ fixed instead of $\wt\m(t)$ and so we need to add a finite term to the renormalized action \eqref{Sren-II} so that 
\be\label{var-CSS}
\d S\sbtx{ren}'=\d\(S\sbtx{ren}+\int \tx dt\;\wt\a\cj^t\wt\m\)=\int \tx dt\(\ct\d\wt\a-\wt\a\co_\j\d\wt\n+\wt\m\d(\wt\a\cj^t)\).
\ee
The subset of PBH transformations that preserves these boundary conditions contains two arbitrary functions of time, namely
\be\label{symmetries-II-CSS}
\ve(t)=\frac{\z(t)}{2\sqrt{LQ}\wt\a} ,\qquad \s(t)=-\frac{\wt L}{2\sqrt{LQ}\wt\a}\z'(t),\qquad \vf(t),
\ee
where $\z(t)$ and $\vf(t)$ are arbitrary. This is a direct consequence of the fact that $Q$ does not transform under PBH transformations and so it does not lead to any constraint on the PBH parameters. Evaluating the Lie brackets \eqref{PBH-commutators} on the BCKVs \eqref{symmetries-II-CSS} gives one copy of the Witt algebra and one copy of a $\Hat u(1)$ Kac-Moody algebra at level zero. Hence, the asymptotic symmetry algebra for constant dilaton solutions with CSS boundary conditions is larger than both that of running dilaton and constant dilaton solutions with Dirichlet boundary conditions.

\paragraph{Conserved charges for CSS boundary conditions}

Evaluating the variation \eqref{var-CSS} of the renormalized action on the BCKVs \eqref{symmetries-II-CSS}, namely
\be
0=\d\sbtx{BCKV} S\sbtx{ren}'=\int \tx dt\(\ct\d\sbtx{BCKV}\wt\a-\wt\a\co_\j\d\sbtx{BCKV}\wt\n+\wt\m\d\sbtx{BCKV}(\wt\a\cj^t)\),
\ee
once again gives trivial conserved charges. In particular, the total derivative term from this variation leads to the conserved charge for local conformal transformations
\be\label{zero-charge-CSS}
\cq[\ve]=\wt\a\ct\ve(t)=0,
\ee
while the fact that $Q$ does not transform under PBH transformations implies that there is no conserved charge associated with the Kac-Moody symmetry. Hence, both the conformal and the Kac-Moody algebras are represented trivial on the Hilbert space of the dual theory with these boundary conditions.

\section{Extended symmetries from 3D embedding}
\label{3Dsymmetries}
\setcounter{equation}{0}

One way to obtain non trivial charges on phase space is by embedding the space of solutions of the 2D EMD model to that of 3D gravity, using the results of section \ref{CFT2}. Although the modes that parameterize the space of solutions remain the same, the 3D embedding allows us to consider a wider class of PBH transformations that involve the Kaluza-Klein circle direction. Applying the analysis at the beginning of the previous section to 3D gravity one finds that 3D PBH transformations act on the modes $g_{(0)_{ij}}$ and $\t_{ij}$ that parameterize the general metric \eqref{3D-metric} as 
\bsub\label{PBH-3D}
\bal
\d\sbtx{PBH}g\sub{0}_{ij}&=\x_o^k\pa_kg_{(0)ij}+g_{(0)kj}\pa_i\x_o^k+g_{(0)ik}\pa_j\x_o^k+\frac{\s}{L}g_{(0)ij},\\
\d\sbtx{PBH}\t_{ij}&=\t_{ik}\pa_j\x_o^k+\t_{jk}\pa_i\x^k+\x^k\pa_k\t_{ij}+\frac{1}{\k_3^2}\(D_{(0)i}D_{(0)j}\s-g_{(0)ij}\square_{(0)}\s\),
\eal
\esub
where $\s(t,z)$ and $\x_o^k(t,z)$ are now arbitrary functions of the 2D boundary coordinates $(t,z)$. Writing $\x_o^t=\ve(t,z)$, $\x_o^z=\vf(t,z)$ and using the identifications between 2D and 3D modes for the two types of solutions given in section \ref{CFT2}, the 3D PBH transformations \eqref{PBH-3D} lead to extended PBH transformations for the 2D modes.  

\subsection{Running dilaton solutions}

In particular, for the running dilaton solutions, \eqref{CFT-metric-I} and \eqref{PBH-3D} give the generalized PBH transformations for the sources
\bsub
\bal\label{PBH-sources-I-3D}
\d\sbtx{PBH}\a &=\pa_t(\ve\a)+\a\s/L-\m\a\pa_z\ve,\\
\d\sbtx{PBH}\b &=\ve\b'+\b\s/L+\b\pa_z(\ve\m+\vf),\\
\d\sbtx{PBH}\m &=\pa_t(\ve\m+\vf)-\m\pa_z(\ve\m+\vf)-\frac{\a^2}{\b^2}\pa_z\ve.
\eal
\esub
Setting the derivatives with respect to the circle direction $z$ to zero gives back the 2D PBH transformations \eqref{PBH-sources-I}. The BCKVs that leave the sources invariant are now determined by the partial differential equations 
\bsub
\bal
&(\pa_t-\m\pa_z)(\m\ve+\vf)-\frac\a\b\pa_z\(\frac\a\b\ve\)=0,\\
&(\pa_t-\m\pa_z)\(\frac\a\b\ve\)-\frac\a\b\pa_z(\m\ve+\vf)=0,\\
&\frac{\s}{L}=\m\pa_z\ve-\frac1\a\pa_t(\ve\a).
\eal
\esub

The general solution of this system of equations involves two arbitrary functions, namely
\be
\frac\a\b\ve=\frac12\(f(z^+)-g(z^-)\),\quad (\m\ve+\vf)=\frac12\(f(z^+)+g(z^-)\),\quad \frac{\s}{L}=-\frac1\b(\pa_+-\pa_-)(\ve\a),
\ee
where $f$ and $g$ are arbitrary functions and
\be\label{z_pm}
z^\pm=z+\int^t\tx dt'\(\m\pm\a/\b\).
\ee
In terms of these coordinates
\be
\pa_z=\pa_++\pa_-,\qquad \pa_t=\(\m+\a/\b\)\pa_++\(\m-\a/\b\)\pa_-,
\ee
while the 2D boundary metric is put into the conformal gauge
\be\label{conf-gauge}
ds\sub{0}^2=\b^2 dz^+dz^-.
\ee
Computing the classical algebra of these asymptotic symmetries one finds two copies of the Witt algebra, one copy for each of the holomorphic and antiholomorphic coordinates. This corresponds to an infinite enhancement of the asymptotic symmetry algebra relative to the pure 2D one, which, as we saw in the previous section, for these boundary conditions contains only two global $u(1)$s.

To construct the conserved charges on the space of solutions, instead of varying the renormalized action with respect to BCKVs as we did in the previous section, we can equivalently use the Ward identities, i.e. the constraints \eqref{2D-constraints}. In terms of the coordinates \eqref{z_pm} these take the form 
\bsub
\label{3D-WIs-I}
\bal
&\pa_-\t_{++}+\b^2\pa_+\(\b^{-2}\t_{+-}\)=0,\\
&\pa_+\t_{--}+\b^2\pa_-\(\b^{-2}\t_{+-}\)=0,\\
&\b^{-2}\t_{+-}=\frac{c}{24\p}R[g_{(0)}]=\frac{c}{3\p}\b^{-4}\(\pa_+\b\pa_-\b-\b\pa_+\pa_-\b\).
\eal
\esub
Notice that since the conformal anomaly does not necessarily vanish in this gauge, the components $\t_{++}$ and $\t_{--}$ are not conserved. However, using the third identity in \eqref{3D-WIs-I}, the first two can be rewritten in the form  
\be
\pa_-\Hat\t_{++}=0,\qquad \pa_+\Hat\t_{--}=0,
\ee
in terms of the modified stress tensor\footnote{This modified stress tensor is conserved, but not fully covariant with respect to 2D boundary diffeomorphisms. However, it corresponds to subtracting the Liouville stress tensor of the Weyl factor of the boundary metric and still leads to well defined charges. We thank Kostas Skenderis and Geoffrey Comp\`ere for useful comments on this point.} 
\be
\Hat\t_{++}=\t_{++}-\frac{c}{12\p}\(\frac{\pa_+^2\b}{\b}-2\(\frac{\pa_+\b}{\b}\)^2\),\qquad \Hat\t_{--}=\t_{--}-\frac{c}{12\p}\(\frac{\pa_-^2\b}{\b}-2\(\frac{\pa_-\b}{\b}\)^2\).
\ee
This allows us to define the conserved charges 
\be\label{3D-charges-I}
\cq_+[\z^+]=\oint \tx dz^+\z^+(z^+)\Hat\t_{++}(z^+),\qquad \cq_-[\z^-]=\oint \tx dz^-\z^-(z^-)\Hat\t_{--}(z^-),
\ee
where
\be\label{3D-BCKVs-I}
\z^+=(\ve\m+\vf)+\frac{\ve\a}{\b}=f(z^+),\qquad \z^-=(\ve\m+\vf)-\frac{\ve\a}{\b}=g(z^-).
\ee

To compute the algebra of these charges we can determine the transformation of the modified stress tensor components $\Hat\t_{++}$ and $\Hat\t_{--}$ by specializing the PBH transformations \eqref{PBH-3D} to the BCKVs \eqref{3D-BCKVs-I}. The result is 
\bsub
\bal
\d\sbtx{BCKV}\Hat\t_{++}&=2\Hat\t_{++}\pa_+\z^++\z^+\pa_+\Hat\t_{++}-\frac{c}{24\p}\pa_+^3\z^+,\\
\d\sbtx{BCKV}\Hat\t_{--}&=2\Hat\t_{--}\pa_-\z^-+\z^-\pa_-\Hat\t_{--}-\frac{c}{24\p}\pa_-^3\z^-,
\eal
\esub
and so these quantities transform as stress tensors with a conformal anomaly corresponding to the Brown-Henneaux central charge. The Dirac brackets
\be
\{\cq_\pm[\z^\pm_1],\cq_\pm[\z^\pm_2]\}=\d_{\z^\pm_1}\cq_\pm[\z^\pm_2],
\ee
therefore lead to two copies of the Virasoro algebra, both with the Brown-Henneaux central charge. However, on the space of running dilaton solutions of the 2D EMD model the stress tensor $\t_{ij}$ takes the specific form \eqref{Liouville-I}. Evaluating the charges \eqref{3D-charges-I} using this expression for the stress tensor we find that all Virasoro generators vanish except for $L_0^\pm$, which correspond to the two global $u(1)$ charges \eqref{2D-charges-I}. For running dilaton solutions with Dirichlet boundary conditions, therefore, although embedding the space of solutions into that of 3D gravity leads to an infinite enhancement of the asymptotic symmetry algebra, but still only a global $u(1)\oplus u(1)$ is represented non trivially on the Hilbert space of the dual theory.

\subsection{Constant dilaton solutions}

Finally let us examine the effect of generalized PBH transformations on the asymptotic symmetries and conserved charges of the constant dilaton solutions. The 2D boundary metric corresponding to constant dilaton solutions is given in eq.  \eqref{CFT-metric-II}. Allowing for an infinitesimal source $\wt\n$ for the irrelevant scalar operator $\co_\j$ corresponds to the slightly more general 2D boundary metric   
\be
ds_{(0)}^2=-2\sqrt{LQ}\wt\a(t)(dt+\wt\m_z(t)dz)(dz+\wt\m(t) dt),
\ee
where $\wt\n\propto\wt\a\wt\m_z$. 

Within the subspace of 3D solutions corresponding to 2D constant dilaton solutions, a generic variation of the renormalized on-shell action takes the form
\be\label{3D-var-II}
\d S\sbtx{ren}=\int \tx d^2x \sqrt{-g_{(0)}}\t^{ij}\d g_{(0)ij}=-2\int \tx d^2x\(\frac{\d\wt\a}{\wt\a}\cp_{tz}+\cp_{zz}\d\wt\m+\cp_{tt}\d\wt\m_z\), 
\ee
where
\be
\cp_{tz}=\t_{tz}-\wt\m\t_{zz},\qquad \cp_{zz}=\t_{zz},\qquad \cp_{tt}=\t_{tt}+2\wt\m^2\t_{zz}-3\wt\m\t_{tz}.
\ee
In terms of these variables and setting $\wt\m_z=0$ the Ward identities \eqref{2D-constraints} become\footnote{Alternatively, these can be obtained by inserting the PBH transformations \eqref{3D-PBH-sources} in the variation \eqref{3D-var-II} of the renormalized on-shell action.}
\be\label{3D-WIs-II}
\cp_{tz}=0,\qquad \pa_z\cp_{tt}=0,\qquad (\pa_t-\wt\m\pa_z)\cp_{zz}=0. 
\ee

The 3D PBH transformations \eqref{PBH-3D} imply that $\wt\m$, $\wt\m_z$ and $\wt\a$ transform respectively as  
\bsub
\label{3D-PBH-sources}
\bal
\d\sbtx{PBH}\wt\m&=\(\pa_t-\wt\m\pa_z\)\(\wt\m\ve+\vf\),\\
\d\sbtx{PBH}\wt\m_z&=\wt\m_z'(\ve+\wt\m_z\vf)+(\pa_z-\wt\m_z\pa_t)(\ve+\wt\m_z\vf),\\
\d\sbtx{PBH}\wt\a&=\pa_t(\ve\wt\a)+\frac{1}{\wt L}\s\wt\a+\wt\a\pa_z(\wt\m\ve+\vf)+\wt\a\wt\m_z\pa_t\vf,
\eal
\esub
while the stress tensor transforms as
\bsub
\label{3D-PBH-stress}
\bal
\d\sbtx{PBH}\t_{tt}&=2\t_{tt}\pa_t\ve+\ve\pa_t\t_{tt}+2\t_{tz}\pa_t\vf+\vf\pa_z\t_{tt}\NO\\
&\hskip0.4cm+\frac{1}{\k_3^2}\(\pa_t^2\s-\wt\m'\pa_z\s-\frac{\wt\a'}{\a}(\pa_t-\wt\m\pa_z)\s-4(\pa_t-\wt\m\pa_z)\pa_z\s\)+\co(\wt\m_z),\\
\d\sbtx{PBH}\t_{zz}&=2\t_{tz}\pa_z\ve+2\t_{zz}\pa_z\vf+\ve\pa_t\t_{zz}+\vf\pa_z\t_{zz}+\frac{1}{\k_3^2}\pa_z^2\s+\co(\wt\m_z),\\
\d\sbtx{PBH}\t_{tz}&=\t_{tt}\pa_z\ve+\t_{zz}\pa_t\vf+\pa_t(\t_{tz}\ve)+\pa_z(\t_{tz}\vf)+\frac{1}{\k_3^2}(2\wt\m\pa_z-\pa_t)\pa_z\s+\co(\wt\m_z).
\eal
\esub
We should point out that in writing these expressions for the generalized PBH transformations of the modes parameterizing the constant dilaton solutions of the 2D theory we have made explicit use of the fact that these modes are only functions of $t$ and not of $z$. These transformations are sufficiently general to describe the symmetries realized on the space of 2D constant dilaton solutions, but in order to obtain the general 3D PBH transformations and to generate these PBH transformations through a Poisson bracket one must allow the modes to depend on both $t$ and $z$.

\paragraph{Asymptotic symmetries for Dirichlet boundary conditions}

The asymptotic symmetries for Dirichlet boundary conditions correspond to PBH transformations that satisfy
\be
\d\sbtx{PBH}\wt\a=0,\qquad \d\sbtx{PBH}\wt\m_z=0,\qquad \d\sbtx{PBH}\wt\m=0.
\ee
The transformations \eqref{3D-PBH-sources} lead to the set of differential equations 
\be
\pa_z\ve=0,\qquad (\pa_t-\wt\m\pa_z)(\wt\m\ve+\vf)=0,\qquad \frac{\s}{\wt L}=-\pa_z\vf-\frac{1}{\wt\a}\pa_t\(\wt\a\ve\),
\ee
whose general solution takes the form
\be\label{3D-BCKVs-II}
\ve(t)=\frac{g(x^+)}{2\wt\a\sqrt{LQ}},\qquad \wt\m\ve+\vf=f\(x^-\),
\ee
where $x^\pm$ are defined in \eqref{liouville-coords} and $f$, $g$ are arbitrary functions. As for the running dilaton solutions, these BCKVs give two copies of the Witt algebra.

\paragraph{Conserved charges for Dirichlet boundary conditions}

In the $x^\pm$ coordinate system Ward identities in \eqref{3D-WIs-II} become respectively
\be
\t_{+-}=0,\qquad \pa_-\t_{++}=0,\qquad \pa_+\t_{--}=0,
\ee 
and hence the charges 
\be
\cq_+[g]=\oint \tx dx^+g(x^+)\t_{++}(x^+),\qquad \cq_-[f]=\oint \tx dx^-f(x^-)\t_{--}(x^-),
\ee
are conserved. Restricting the PBH transformations \eqref{3D-PBH-stress} to the BCKVs \eqref{3D-BCKVs-II} we find that $\t_{\pm\pm}$ transform as 
\bsub
\bal
\d\sbtx{BCKV}\t_{++}&=2\t_{++}\pa_+g+g\pa_+\t_{++}-\frac{c}{24\p}\pa_+^3g,\\
\d\sbtx{BCKV}\t_{--}&=2\t_{--}\pa_-f+f\pa_-\t_{--}-\frac{c}{24\p}\pa_-^3f.
\eal
\esub
It follows that the two copies of the Witt algebra realized on the BCKVs turn into two copies of the Virasoro algebra on the space of solutions, both with the Brown-Henneaux central charge. However, the explicit form of the stress tensor for constant dilaton solutions given in \eqref{Liouville-tensor-II} and \eqref{Liouville-sol-II} implies that only the charges $\cq_+$ are realized non trivially on the phase space of 2D constant dilaton gravity, while only a $u(1)$ survives from the other copy.

\paragraph{Asymptotic symmetries for CSS boundary conditions} In order to impose CSS boundary conditions we need to add a boundary term to the renormalized action that implements the appropriate Legendre transformation, namely  
\be
S\sbtx{ren}\to S\sbtx{ren}+2\int \tx d^2x\;(1-\wt\m\wt\m_z)\wt\m\t_{zz}.
\ee
Note that the $\co(\wt\m_z)$ term affects the form of the resulting canonical variables and it is necessary to get consistent results. A generic variation of the resulting action takes the form  
\be\label{3D-var-CSS}
\d \Big(S\sbtx{ren}+2\int \tx d^2x\;(1-\wt\m\wt\m_z)\wt\m\t_{zz}\Big)=-2\int \tx d^2x\(\frac{\d\wt\a}{\wt\a}\cp_{tz}-\wt\m\d\cp_{zz}+(\cp_{tt}+\wt\m^2\cp_{zz})\d\wt\m_z\), 
\ee
where again
\be
\cp_{tz}=\t_{tz}-\wt\m\t_{zz},\qquad \cp_{zz}=\t_{zz},\qquad \cp_{tt}=\t_{tt}+2\wt\m^2\t_{zz}-3\wt\m\t_{tz}.
\ee
Demanding invariance of this action under the generic PBH transformations \eqref{3D-PBH-sources} and \eqref{3D-PBH-stress} implies the stronger Ward identities (cf. \eqref{3D-WIs-II}) 
\be
\pa_z\cp_{tt}=0,\qquad \pa_t\cp_{zz}=\pa_z\cp_{zz}=0,
\ee
and so $\cp_{zz}=\t_{zz}$ must be constant for CSS boundary conditions. 

The asymptotic symmetries for CSS boundary conditions are obtained from the conditions
\be
\d\sbtx{PBH}\wt\a=0,\qquad \d\sbtx{PBH}\wt\m_z=0,\qquad \d\sbtx{PBH}\cp_{zz}=0, 
\ee
which translate to the differential equations 
\be
\pa_z\ve=0,\qquad \Big(2\t_{zz}\pa_z-\frac{\wt L}{\k_3^2}\pa_z^3\Big)\vf=0,\qquad \frac{\s}{\wt L}=-\pa_z\vf-\frac{1}{\wt\a}\pa_t\(\wt\a\ve\). 
\ee
The general solution contains four arbitrary functions and in terms of the variables $x^+$ (defined in \eqref{liouville-coords} ) and $z$ takes the form
\be\label{3D-BCKVs-II-CSS}
\ve=\frac{\z(x^+)}{2\wt\a\sqrt{LQ}},\qquad \vf=\vf_o(x^+)+a_+(x^+)e^{2\sqrt{\frac{Q}{L}}z}+a_-(x^+)e^{-2\sqrt{\frac{Q}{L}}z},
\ee
where $\ve(x^+)$, $\vf_o(x^+)$ and $a_\pm(x^+)$ are arbitrary functions. The algebra these BCKVs generate is one copy of the Witt algebra as well as an $\Hat{sl}(2,\bb R)$ Kac-Moody algebra at level zero. This is precisely the asymptotic symmetry algebra with CSS boundary conditions found in \cite{Avery:2013dja}. However, to get non trivial charges for the full $\Hat{sl}(2,\bb R)$ it is necessary that $\wt\m$ depends on both $t$ and $z$ (see e.g. eq. (2.13) in \cite{Avery:2013dja}), which takes us outside the space of solutions of the 2D theory. Hence, only a $\Hat{ u}(1)\subset \Hat{sl}(2,\bb R)$ subalgebra is realized non trivially on the space of constant dilaton solutions in 2D, which corresponds to the original symmetry algebra in \cite{Compere:2013bya}.

\paragraph{Conserved charges for CSS boundary conditions} Inserting the BCKVs \eqref{3D-BCKVs-II-CSS} (with $a_\pm=0$) in the variation \eqref{3D-var-CSS} of the on-shell action and keeping the total derivative terms leads to the conserved charges   
\be
\cq_+[\ve]=\int_0^{2\p } \hskip-0.2cm\tx d\f\(\t_{++}-\frac{2\t_{zz}\wt\m^2}{(2\sqrt{LQ}\wt\a)^2}\)\z(x^+),\qquad \cq_-[\vf_o]=-\int_0^{2\p} \hskip-0.2cm\tx d\f\;\frac{2\t_{zz}\wt\m}{2\sqrt{LQ}\wt\a}\vf_o(x^+),
\ee
where $x^+=\bar t+\f$, $z=\bar t-\f$, and $\f\sim \f+2\p$. The algebra these charges generate can be computed by specializing the transformations \eqref{3D-PBH-sources} and \eqref{3D-PBH-stress} to the BCKVs \eqref{3D-BCKVs-II-CSS} with $a_\pm=0$, namely
\bal
\d\sbtx{BCKV}\t_{++}&=2\t_{++}\pa_+\z+\z\pa_+\t_{++}+\frac{2\t_{zz}\wt\m}{2\sqrt{LQ}\wt\a}\pa_+\vf_o-\frac{c}{24\p}\pa_+^3\z,\\
\d\sbtx{BCKV}\(\frac{\wt\m}{2\sqrt{LQ}\wt\a}\) &=\pa_+\(\frac{\wt\m}{2\sqrt{LQ}\wt\a}\z+\vf_o\).
\eal
Defining the generators 
\be
L_n=\cq_+[e^{-inx^+}],\qquad J_n=\cq_-[e^{-inx^+}],\qquad n\in\bb Z, 
\ee
we obtain the algebra
\bal
i\{L_m,L_n\}&=(m-n)L_{m+n}+\frac{c}{12}m^3\d_{n+m,0},\\
i\{L_m,J_n^{0,\pm}\}&=-nJ_{m+n}^{0,\pm},\\
i\{J_m^0,J_n^0\}&=-4\p \t_{zz}m\d_{m+n,0}=-2\D m\d_{m+n,0},
\eal
in agreement with \cite{Compere:2013bya}.

\section*{Acknowledgments}

We would like to thank Jan de Boer, Alejandra Castro, Geoffrey Comp\`ere, Daniel Grumiller and Finn Larsen for enlightening discussions and email correspondence, as well as Ok Song An for collaboration at the early stages of this project. We are especially grateful to Alejandra Castro for very useful comments on the manuscript. We thank NORDITA and CERN for hospitality and partial financial support during the completion of this project. M.C. would like to thank the Beijing Normal University for hospitality during the course of the work. This research is supported in part by the DOE Grant Award DE-SC0013528, (M.C.), the Fay R. and Eugene L. Langberg Endowed Chair (M.C.) and the Slovenian Research Agency (ARRS) (M.C.).

\appendix

\renewcommand{\thesection}{\Alph{section}}
\renewcommand{\theequation}{\Alph{section}.\arabic{equation}}


\section{4D subtracted geometries and Kaluza-Klein Ans\"atze}
\label{appendix}
\setcounter{equation}{0}

In this appendix we summarize the four-dimensional STU model and its subtracted geometries in the parameterization introduced in \cite{An:2016fzu}. Moreover, we provide the Kaluza-Klein Ansatz for the reduction of the general rotating subtracted geometries to two dimensions.

The general rotating subtracted geometries with three equal magnetic charges and one electric charge are solutions of the action \cite{Cvetic:2012tr,Virmani:2012kw}
\bal\label{STU-Action-Reduced}
S = & \frac{1}{2\k_4^2}\int_\cm \text{d}^4\mathbf{x}\sqrt{-g}\Big(R[g]-\frac 32\pa_\mu\h\pa^\mu\h-\frac32 e^{2\h}\pa_\mu\c\pa^\mu\c-\frac14 e^{-3\h}F_{\m\n}^0F^{0\m\n}\NO\\
& \hskip0.5in -\frac34 e^{-\h}( F+\c F^0)_{\m\n}( F+\c F^0)^{\m\n}\Big)\NO\\
&  -\frac{1}{8\k_4^2}\int_{\cm}\text{d}^4\mathbf{x}\sqrt{-g}\e^{\m\n\r\s}\(\c^3F^0_{\m\n}F^0_{\r\s}+3\c^2F^0_{\m\n} F_{\r\s}+3\c  F_{\m\n} F_{\r\s} \),
\eal 
and take the form
\bal
\label{subtracted-simple}
e^\h&=\frac{B^2/\ell^2}{\sqrt{r+\ell^2\o^2\sin^2\th}},\qquad \c=\frac{\ell^3\o}{B^2}\cos\th,\NO\\
A^0&=\frac{B^3/\ell^3}{r+\ell^2\o^2\sin^2\th}\(\sqrt{r_{+}r_{-}}\;kdt+\ell^2\o\sin^2\th d\phi\),\NO\\
A&=\frac{B\cos\th}{r+\ell^2\o^2\sin^2\th}\(-\o\sqrt{r_{+}r_{-}}\;kdt+rd\phi\),\NO\\
ds^2&=\sqrt{r+\ell^2\o^2\sin^2\th}\(\frac{\ell^2dr^2}{(r-r_{-})(r-r_{+})}-\frac{(r-r_{-})(r-r_{+})}{r}k^2dt^2+\ell^2 d\th^2\)\NO\\
&\hskip2.6in+\frac{\ell^2 r\sin^2\th}{\sqrt{r+\ell^2\o^2\sin^2\th}}\(d\f-\frac{\o\sqrt{r_+ r_-}}{r}k dt\)^2.
\eal
Note that the magnetic charge $B$ must be non zero for these solutions to be related to solutions of the two-dimensional model \eqref{action}. The coordinate transformations and map of parameters that relate this form of the solutions to that found in previous works are given in \cite{An:2016fzu}. The form \eqref{subtracted-simple} of the subtracted geometries makes it manifest that they are asymptotically conformal to AdS$_2\times S^2$, which implies that holography for these black holes requires a Kaluza-Klein reduction on the internal $S^2$.

The EMD theory \eqref{action} can be obtained from the STU model \eqref{STU-Action-Reduced} by means of the consistent Kaluza-Klein Ansatz
\bsub\label{KK-relations}
\bal
e^{-2\h}&=e^{-2\j}+\l^2B^2\sin^2\th,\qquad\c=\l B\cos\th,\\
e^{-2\h}A^0&=e^{-2\j}A^{(2)}+\l B^2\sin^2\th d\f,\qquad
A+\c A^0=B\cos\th d\f,\\
e^\h ds_4^2&=ds_2^2+B^2\(d\th^2+\frac{\sin^2\th}{1+\l^2B^2e^{2\j}\sin^2\th}(d\f-\l A^{(2)})^2\),
\eal
\esub
with the identifications
\be
L=2B,\qquad \k_2^2=\k_4^2/\p L^2.
\ee
Moreover, the constant $\l$ that parameterizes the family \eqref{KK-relations} of Kaluza-Klein Ans\"atze is identified with the rotational parameter of the rotating black hole solutions \eqref{subtracted-simple} as
\be
\l =\o\ell^3/B^3.
\ee 
This parameter must be a fixed constant for this Ansatz to be well defined. Since $\l$ does not enter as a parameter of the 2D model, it can be viewed as a modulus of the dual theory. Moreover, any given solution in two dimensions can be uplifted to different solutions of the STU model \eqref{STU-Action-Reduced}, with different values of $\l$. One of the advantages of the parameterization \eqref{subtracted-simple} of the subtracted geometries is that it clearly separates the various parameters of the 4D solution into a) parameters of the 2D theory ($B$) b) integration constants of the two-dimensional theory ($k$, $\ell$, and $r_\pm$), and c) parameters that characterize the Kaluza-Klein Ansatz but do not survive in the 2D theory ($\l$). This separation of the different types of parameters is crucial for understanding holography for such black holes.  

The easiest way to show that the Kaluza-Klein Ansatz \eqref{KK-relations} is consistent is to first uplift the STU model to five dimensions \cite{Cvetic:2012tr,Virmani:2012kw}, and then reduce on the internal $S^2$, as is indicated with the blue arrows in Fig. \ref{fig1}. The uplift to fine dimensions is given by
\bsub\label{KK-Ansatz}
\bal
ds_5^2&=e^\h ds_4^2+e^{-2\h}(dz+A^0)^2,\\
&=ds^2_3+B^2d\O_\l^2\NO\\
&=ds_2^2+e^{-2\j}(dz+A^{(2)})^2+B^2d\O_\l^2\NO\\
A^{(5)}&=B\cos\th\(d\f+\l dz\),
\eal
\esub
where
\be
d\O_\l^2=d\th^2+\sin^2\th\(d\f+\l dz\)^2,
\ee
and so the coordinate $z$ must be periodic with period $R_z=2\p/\l$.

\section{Comparison with \cite{Castro:2014ima} and \cite{Hartman:2008dq}}
\label{appendixB}
\setcounter{equation}{0}

In this appendix we briefly comment on the relation of our constant dilaton results with those in \cite{Castro:2014ima} and \cite{Hartman:2008dq}. Starting with \cite{Castro:2014ima}, there are two main differences, which lead to somewhat different conclusions. The first is the boundary counterterms used in \cite{Castro:2014ima}, given in eq. (A.22), which agree (taking into account the different normalization of the gauge field and modulo an overall factor of the AdS radius in \cite{Castro:2014ima} that we believe is a typo) with the boundary counterterms \eqref{counterterms-1}. However, as we argued in section \ref{dict}, the correct boundary terms for constant dilaton solutions are instead of the form \eqref{exotic-bt} and \eqref{ct-constant}. Furthermore, in addition to the CSS boundary conditions considered in \cite{Castro:2014ima}, we also consider Dirichlet boundary conditions on the gauge field for constant dilaton solutions. 

Even though the asymptotic symmetries obtained in section A.2 of \cite{Castro:2014ima} coincide with our result for CSS boundary conditions in \eqref{symmetries-II-CSS}, the use of different boundary counterterms in \cite{Castro:2014ima}, which leads to a different identification of the dual operators, prevents a quantitative comparison of the subsequent analyses. A number of qualitative observations can be made, however. Firstly, for a given choice of boundary terms, there are three distinct variations of the on-shell action that contain distinct physical information. Varying the action with respect to arbitrary sources one obtains the conjugate operators. Varying the action with respect to generic PBH transformations gives the corresponding Ward identities. As we have seen, the Ward identities are the generators of generic PBH transformations through the Poisson bracket. Finally, the variation of the action with respect to the asymptotic symmetries, i.e. PBH transformations that leave the sources invariant, leads to the corresponding conserved charges. In \cite{Castro:2014ima} these three distinct variations lead respectively to the quantities in (A.15) (operators), (A.17) (conserved current) and (A.21) or (A.23) (conserved charges or generators of asymptotic symmetries). 

However, the corresponding quantities with our boundary terms are different. For constant dilaton solutions we have found that in the case of Dirichlet boundary conditions on the gauge field the dual operators are as given in 
\eqref{1pt-fns-const}. For CSS boundary conditions the only change is that there is no local current operator but only the non local `Polyakov loop' operator $\int \tx dt\;\wt\m(t)$. In particular, the stress tensor vanishes identically. The corresponding Ward identities are given in \eqref{WIs-II}, which are trivial at zero source for the irrelevant scalar operator $\co_\j$. Moreover, there is no charge conservation constraint for CSS boundary conditions since $Q$ does not transform under (2D) PBH transformations. As a result, there is no associated U(1) charge, while the conserved charges for conformal transformations in \eqref{zero-charge-CSS} vanish identically due to the vanishing of the stress tensor. Besides the asymptotic symmetries, this last conclusion is the only point where our analysis agrees with that of appendix A in \cite{Castro:2014ima}. In particular, we indeed find that the (true time, boundary) Hamiltonian, i.e. $\ct$, vanishes identically in this theory and the only non trivial observable is the VEV of the irrelevant scalar operator $\co_\j$. As we saw in \eqref{anomalous-scalar}, this operator does transform anomalously under local conformal transformations and so the corresponding central charge can be associated with the number of degrees of freedom in this theory.       

Finally, our analysis does not contradict the results of \cite{Hartman:2008dq}, nor the improvements made in appendix B of \cite{Castro:2014ima}. However, we do not see a direct connection between the two analyses. The calculation in  \cite{Hartman:2008dq} is a bulk calculation treating the 2D theory in conformal gauge as a perturbative 2D CFT on the strip. In particular, the Virasoro and local U(1) constraints considered there are directly analogous to our first class constraints \eqref{constraints}, although the choice of gauge is different (conformal versus Fefferman-Graham). The local transformations generated through the Poisson bracket by these constraints is the full set of PBH transformations, whose analogue in the conformal gauge is holomorphic and antiholomorphic conformal and U(1) gauge transformations. However, the classical Poisson bracket of these constraints does not contain anomalous terms. In the standard Brown-Henneaux analysis, non trivial central charges are generated by considering the subset of PBH transformations that correspond to the asymptotic symmetries and imposing the constraints strongly. The generating functions of local transformations then vanish, except for boundary terms that correspond to the conserved charges. It is in the Dirac brackets of these conserved charges that the non trivial central terms appear. However, we have shown explicitly that these charges are identically zero. The Dirac brackets that \cite{Hartman:2008dq} consider instead are those of the constraints in the quantized bulk theory. Moreover, the argument given there provides a relation between the Virasoro central charge for the twisted stress tensor dictated by the boundary conditions (the central charge of the untwisted stress tensor must be zero for a consistent theory of quantum gravity \cite{Strominger:1998yg}) and the level of the $u(1)$ Kac-Moody current algebra, but the Kac-Moody level itself is not derived. 

We believe that a better understanding of the isomorphism between the bulk CFT$_2$ Hilbert space and that of the dual CFT$_1$ is required in order to clarify how the symmetry algebras act and what is the relation between the representations carried by each Hilbert space. The qualitative similarities between the bulk symmetry algebra discussed in \cite{Hartman:2008dq} and that of the boundary theory once the extra circle direction is taken into account, as observed in \cite{Castro:2014ima}, might indicate that the full Kaluza-Klein tower of modes in AdS$_2$ must be considered in order to be able to match the bulk and boundary Hilbert spaces and the representations of the symmetry algebras.



\bibliographystyle{jhepcap}
\bibliography{ads2}

\end{document}

%% file: macros.tex
\def\){\right)}
\def\({\left( }
\def\]{\right] }
\def\[{\left[ }

\def\NO{\nonumber}

\newcommand{\be}{\begin{equation}}
\newcommand{\ee}{\end{equation}}

\def\bea{\begin{eqnarray}}
\def\eea{\end{eqnarray}}

\def\bal#1\eal{\begin{align}#1\end{align}}

\def\bald{\begin{aligned}}
\def\eald{\end{aligned}}

\def\bsub{\begin{subequations}}
\def\esub{\end{subequations}}

\def\beqx{\begin{displaymath}}
\def\eeqx{\end{displaymath}}

\newcommand{\bmat}{\left(\begin{array}}
\newcommand{\emat}{\end{array}\right)}

\def\a{\alpha}
\def\b{\beta}
\def\c{\chi}
\def\d{\delta}
\def\e{\epsilon}
\def\f{\phi}
\def\g{\gamma}
\def\h{\eta}
\def\j{\psi}
\def\k{\kappa}
\def\l{\lambda}
\def\m{\mu}
\def\n{\nu}
\def\o{\omega}
    
\def\p{\pi}

    \def\th{\theta}
\def\r{\rho}
\def\s{\sigma}
\def\t{\tau}
\def\x{\xi}
\def\z{\zeta}
\def\D{\Delta}

\def\L{\Lambda}
\def\O{\Omega}
    
\def\P{\Pi}

\def\ve{\varepsilon}

\def\vf{\varphi}

\def\ca{{\cal A}}

\def\cc{{\cal C}}
\def\cd{{\cal D}}

\def\cf{{\cal F}}
\def\cg{{\cal G}}
\def\ch{{\cal H}}

\def\cj{{\cal J}}

\def\cl{{\cal L}}
\def\cm{{\cal M}}
\def\cn{{\cal N}}
\def\co{{\cal O}}
\def\cp{{\cal P}}
\def\cq{{\cal Q}}

\def\cs{{\cal S}}
\def\ct{{\cal T}}

\def\bb#1{\ensuremath{\mathbb{#1}}} 

\def\bo{{\raise-.3ex\hbox{\large$\Box$}}}               
\def\pa{\partial}                                       
\def\face{{\raise.2ex\hbox{$\displaystyle \bigodot$}\mskip-2.2mu \llap {$\ddot
        \smile$}}}                                   
\def\>{\rangle}                                      
\def\<{\langle}                                      

\def\tx#1{\text{#1}}
\def\sbtx#1{{}_{\rm #1}}                           
\newcommand{\sub}[1]{\phantom{}_{(#1)}\phantom{}}    
\def\wt#1{\widetilde{#1}}                            
\def\Hat#1{\widehat{#1}}                             
\def\leftrightarrowfill{$\mathsurround=0pt \mathord\leftarrow \mkern-6mu
        \cleaders\hbox{$\mkern-2mu \mathord- \mkern-2mu$}\hfill
        \mkern-6mu \mathord\rightarrow$}        
\def\dvec#1{\vbox{\ialign{##\crcr
        \leftrightarrowfill\crcr\noalign{\kern-1pt\nointerlineskip}
        $\hfil\displaystyle{#1}\hfil$\crcr}}}           

\def\-{\hphantom{-}}

